\documentclass[fleqn,usenatbib]{mnras}

\usepackage{newtxtext,newtxmath}

\usepackage[T1]{fontenc}

\DeclareRobustCommand{\VAN}[3]{#2}
\let\VANthebibliography\thebibliography
\def\thebibliography{\DeclareRobustCommand{\VAN}[3]{##3}\VANthebibliography}

\usepackage[flushleft]{threeparttable}
\usepackage[dvipsnames]{xcolor}
\usepackage{graphicx}	
\usepackage{amsmath}	

\definecolor{mypink1}{rgb}{0.858, 0.188, 0.478}
\definecolor{mygreen}{rgb}{0, 0.5, 0}

\newcommand{\Msun}{{\rm M_{\odot}}}
\newcommand{\Gyr}{{\rm Gyr}}

\newcommand{\kpc}{{\rm kpc}}
\newcommand{\kms}{{\rm km\cdot s^{-1}}}

\newcommand{\td}{t_{\rm d}}
\newcommand{\tdiss}{t_{\rm diss}}

\newcommand{\tacc}{t_{\rm turb,acc}}

\newcommand{\epsff}{\epsilon_{\rm ff}}
\newcommand{\Rd}{R_{\rm d}}
\newcommand{\Vd}{V_{\rm d}}
\newcommand{\Mg}{M_{\rm g}}
\newcommand{\Qg}{Q_{\rm g}}
\newcommand{\pms}{\left(\frac{p}{m}\right)_*}
\newcommand{\lbrac}[1]{\left(#1\right)}

\title[Turbulence in discs]{The evolution of turbulent galactic discs: gravitational instability, feedback and accretion}

\author[O. Ginzburg et al.]{
Omri Ginzburg,$^{1}$\thanks{E-mail: omry.ginzburg@mail.huji.ac.il}
Avishal Dekel$^{1,2}$
Nir Mandelker$^{1}$ and Mark R. Krumholz$^{3,4}$
\\
$^{1}$Racah Institute of Physics, The Hebrew University, Jerusalem 91904 Israel\\
$^{2}$SCIPP, University of California, Santa Cruz, CA 95064, USA\\
$^{3}$Research School of Astronomy and Astrophysics, Australian National University, Canberra, ACT 2611, Australia\\
$^{4}$Australian Research Council Centre of Excellence for All Sky Astrophysics in 3 Dimensions (ASTRO 3D), Australia
}

\date{Accepted XXX. Received YYY; in original form ZZZ}

\pubyear{2021}

\begin{document}
\label{firstpage}
\pagerange{\pageref{firstpage}--\pageref{lastpage}}
\maketitle

\begin{abstract}
We study the driving of turbulence in star-froming disc galaxies of different masses at different epochs, using an analytic "bathtub" model.
The disc of gas and stars is assumed to be in marginal Toomre instability. 
Turbulence is assumed to be sustained via an energy balance between its dissipation and three simultaneous energy sources.
These are stellar feedback, inward transport due to disc instability and clumpy accretion via streams. 
The transport rate is computed with two different formalisms, with similar results.
To achieve the energy balance, the disc self-regulates either the mass fraction in clumps or the turbulent viscous torque parameter.
In this version of the model, the efficiency by which the stream kinetic energy is converted into turbulence is a free parameter, $\xi_a$. We find that the contributions of the three energy sources are in the same ball park, within a factor of $\sim\!2$ in all discs at all times. In haloes that evolve to a mass $\leq 10^{12}\,\Msun$ by $z=0$ ($\leq 10^{11.5}\,\Msun$ at $z\!\sim\!2$), feedback is the main driver throughout their lifetimes. Above this mass, the main driver is either transport or accretion for very low or very high values of $\xi_a$, respectively. For an assumed $\xi_a(t)$ that declines in time, galaxies in halos with present-day mass $>\!10^{12}$ M$_\odot$ make a transition from accretion to transport dominance at intermediate redshifts, $z\! \sim\!3$, when their mass was $\geq\!10^{11.5}\,\Msun$. The predicted relation between star-formation rate and gas velocity dispersion is consistent with observations.
\end{abstract}




\section{Introduction}
Most of the star formation in galaxies tends to occur in galactic discs \citep{Wuyts11,HC15}. This observational result makes the evolution of gaseous discs a key for our understanding of galaxies.

\smallskip According to our current knowledge, very thin discs, with rotation-to-dispersion velocity ratios $\Vd/\sigma_{\rm g}\!>\!5$, are expected to emerge at or after $z\!\sim\!1$ \citep{Kassin12}. At higher redshifts, as a result of intense accretion through streams and associated dynamical instabilities, discs tend to be more perturbed, with $\Vd/\sigma_g\sim3$ \citep{Genzel2006,Genzel08,Simons2017,Forster2018}, although some observations suggest that massive cold discs can form as early as redshift $z\!\approx\!4$, with $\Vd/\sigma_{\rm g}\!\sim\!10$, at least for short periods \citep{Rizzo20,Neelman20,Lelli21}. Theoretically, \cite{DekelDiscs} found, using analytical estimates as well as cosmological simulations, that frequent spin flips, driven by mergers, tend to destroy discs within an orbital time when they reside in haloes with mass ${M_{\rm h}\lesssim\!2\cdot10^{11}\,\Msun}$, roughly independent of redshift. \cite{Kretschmer21} found cold discs with $\Vd/\sigma_{\rm g}\!\sim\!5$ in cosmological simulations at redshift $z\!\sim\!3.5$, with an even colder molecular disc, of $\Vd/\sigma_{\rm g,mol}\!\sim\!8$.

\smallskip Self-gravitating, rotating discs tend to undergo various instabilities, collectively termed violent disc instabilities (VDI). The most common of these instabilities is the Toomre instability \citep{Toomre64, Noguchi99, DekelSariCeverino}, governed by the Toomre-$Q$ parameter (see \S\ref{sec:disc_instability}), which expresses the balance between self gravity on one hand, and pressure and rotation on the other hand. A $Q$ parameter of around unity indicates marginal equilibrium between these forces.
Observed galaxies, both local and at high redshifts, indeed seem to have $Q$ parameter of order unity \citep{Genzel11,RF13,Genzel14,Obreschkow15}. 

\smallskip In order to maintain a constant level of $Q$ near unity, for a given rotation velocity, the turbulence at the disc must remain at the level required to counteract self-gravity without overstabilising the disc. However, turbulence, especially at supersonic levels as observed over a wide range of redshifts \citep{Kassin12,Ianjamasimanana15,Wisnioski15,Stott16} tends to decay rapidly in a dynamical time. Hence, in order to sustain the turbulence levels required for marginal Toomre instability, a continuous supply of energy to turbulence is required. It is therefore important to understand the different energy inputs from various mechanisms that drive turbulence in marginally Toomre-unstable discs.

\smallskip There are several physical processes that occur regularly in discs that can provide a continuous supply of energy to turbulence. First, star formation results in supernova explosions and various forms of stellar feedback. These release large amounts of energy and momentum to their surroundings \citep{Draine11}. A fraction of this energy turns into heat, while the rest can turn into kinetic energy, partially driving outflows and partially driving turbulence \citep{HH16}. 

A second energy source for turbulence is radial mass transport. When mass is transported inwards through the disc and down the overall potential well, there is a gravitational energy gain that could be converted into turbulence \citep{Wada02}. A natural driver of inward mass transport is the disc instability itself, where the deviations from circular symmetry exert torques that tend to drive angular momentum out, and therefore transport mass in by conservation of angular momentum. As previously mentioned, the instability tends to self-regulate and maintain $Q\sim1$. The self-regulation mechanism can be understood as follows: If $Q$ falls below unity, the torques get stronger, the inward transport rate becomes higher, the gravitational energy gain by this transport down the potential well enhances the turbulence, which in turn increases $Q$.
If $Q$ becomes larger than unity,  the torques weaken, the transport rate is reduced, the energy gain becomes smaller, the turbulence is reduced and $Q$ becomes smaller. While the above picture gives an idea of how disc-instability can give rise to mass transport and turbulence in a self-regulated way, the precise details of how VDI generates inwards mass transport are unclear. We here consider two possibilities, detailed below.

 \smallskip The first mechanism is encounters between giant clumps in the disc. Galaxies, in the redshift range $z\approx 1-3$, are typically observed to host multiple massive, star forming giant clumps \citep{Elmegreen05,Genzel11,Shibuya2016,G18,HC20,Ginzburg21}. These clumps are consistently found in UV, ${\rm H\alpha}$ and CO observations \citep{Forster2009,Forster2011,Dessauges2019}, which serve as tracers of star formation, as well as in the optical regime \citep{HC20}. Giant star forming clumps are also observed, though less frequently, in local, gas rich systems \citep{Fisher17,Lenkic21}. The emergence of these clumps is a robust theoretical prediction, found in isolated as well as cosmological simulations \citep{Noguchi99,Agretz09,DekelSariCeverino,Genel12b,Bournaud2014,M14,Oklopcic2017,M17}. The rate at which clumps interact with each other and their surroundings, along with the resulting torques, can be computed in the Toomre framework to get an estimate for the mass transport inwards \citep{DekelSariCeverino}. This transport of mass is partially due to clump migration and partially to interclump inflow.

\smallskip The second mechanism that we consider below for inducing mass transport is turbulent viscosity. Turbulence can be modeled as an effective viscosity that is added to the gas, well beyond the molecular viscosity. We use the standard theory for viscous accretion discs by \cite{Shakura73}, followed by \cite{Balbus99} and simulations by \cite{Gammie}. \cite{KB10} introduced a hydrodynamical model for turbulent viscous discs that includes the effect of star formation. They showed that the systems usually reach a steady state, which allows one to compute the torques induced by this effective viscosity, and from that the resulting mass transport inwards.

\smallskip Finally, the third energy source we consider for driving turbulence in the disc is the intense external accretion onto it. The primary mechanism by which high-redshift galaxies obtain fresh gas is via accretion along cold streams \citep{Keres05,DB06,Dekel09}.Typically, the disc is fed by three streams which are confined to a plane with most of the inflowing mass co-rotating with the disc \citep{Danovich12,Danovich15}. As the streams flow into the disc, they are subject to various instabilities, including gravitational instabilities, Kelvin-Helmholtz instability and thermal instabilities \citep{M16,Padnos18,Aung19,M18_GC,M19,M20_cooling}, which cause the streams to fragment and form bound clumps within them \citep{M18_GC,M20_cooling}. Similar instabilities also occur in ISM filaments inside GMCs \citep{Clarke15,Clarke16,Clarke17}. Such clumpy accretion can be very efficient at transferring the kinetic energy associated with the accretion into turbulent energy within the disc, when the clumps collide with the disc \citep{KH10}.

\smallskip The nature of turbulent support in discs undergoing VDI has been studied with various levels of complexity. \cite{DekelSariCeverino} and \cite{Cacciato12} studied the evolution of discs in which clump encounters are the main drivers of turbulence and mass transport, without the input of supernova, stellar feedback or accretion. \cite{KB10} and \cite{Forbes12,Forbes14} studied the effects of disc instability on the turbulence within the disc, using the turbulent viscosity formalism. \cite{FG13} and \cite{HH16} argued that supernova feedback driven turbulence is enough to explain the high levels of turbulence in galaxies. However, this requires a very high star formation rate, with a star formation law $\dot{\Sigma}_{\rm SF}\propto \Sigma_g^2$, where $\dot{\Sigma}_{\rm SF}$ and $\Sigma_g$ are the star fromation rate and gas densities, steeper than observed \citep{Kennicutt98}. {\cite{ElmegreenBurkert10} studied the driving of turbulence by accretion in young discs, finding that accretion alone cannot sustain turbulence at high levels for long periods of time. However, their considered efficiency with which the accretion drives turbulence was low, and does not reflect the possible enhancement of turbulence driving by the clumps in the streams.}  \cite{Genel12} considered the energy input by accretion and disc instability, but they did not include the input by supernova or stellar feedback, and their treatment of mass transport due to disc instability did not take into account the turbulent state of the disc. 

\smallskip \cite{K18} developed a comprehensive model for the evolution of disc galaxies,accounting for turbulence driven by both supernova feedback and transport, extending the framework of turbulent viscous torques developed in \cite{KB10}, but without including the independent contribution of accretion to the driving of turbulence, but only as a source of mass to the disc. They found that at high redshifts, turbulence in galaxies tends to be driven primarily by transport, while supernova feedback is sufficient only in low mass galaxies at low redshifts.

\smallskip In this paper, inspired by the recent studies of \cite{M18_GC}, \cite{Aung19} and \cite{M20} regarding the clumpiness of cosmological streams, we consider simultaneously the three energy sources for turbulence. In particular, we extend the earlier studies to address the role of accretion-driven turbulence in sustaining turbulence in discs as a function of halo mass and redshift. We wish to ascertain how different prescriptions for mass transport and for the efficiency of converting accretion energy into turbulence affect the result. We defer a study of how exactly the different mechanisms drive turbulence to future work, and focus here only on their energy input to turbulence. 

The paper is organized as follows. In \S\ref{sec:the_model}, we present our model. To connect to earlier studies, we model several basic properties of the galaxies using an approach similar to that of \cite{K18} while adding the necessary new prescriptions for transport and accretion. In \S\ref{sec:mass_budget} we describe the different mass sources and sinks, and in \S\ref{sec:turb_budget} we address the different drivers of turbulence and their corresponding energy input. In \S\ref{sec:Results}, we show the results of numerically integrating the mass conservation equation, and study the relative roles of the different energy sources in driving disc turbulence. In \S\ref{sec:discussion}, we discuss caveats of the model. We present our conclusions in \S\ref{sec:conclusions}.
\section{The Model}\label{sec:the_model}
We model the galaxy as a disc embedded in a dark matter halo. The discs undergo accretion of fresh gas from the cosmic web, and can lose gas by either star formation or mass transport from the disc to a central bulge, which we consider as separate from the disc. Our model is based on two main assumptions:
\begin{itemize}
    \item The disc maintains marginal Toomre instability, with a two component Toomre-$Q$ around unity (\S\ref{sec:disc_instability}),
    \item The disc maintains turbulent energy balance, where turbulent dissipation is balanced by driving via the three sources discussed above (\S\ref{sec:QE_turb}).
\end{itemize}
In the following sections, we lay out in detail each component of our model.
\subsection{Galaxy-halo connection}\label{sec:gal_halo_con}
Galaxies are assumed to be embedded in dark matter haloes with virial radii and velocities determined by cosmology, given approximately by \citep{Dekel13}
\begin{equation}
    \frac{R_{\rm v}}{100\,\kpc} \approx 0.54 M_{\rm h,12}^{1/3}\lbrac{\frac{1+z}{3}}^{-1},
\end{equation}
\begin{equation}
\frac{V_{\rm v}}{200\,\kms}\approx M_{\rm h,12}^{1/3}\lbrac{\frac{1+z}{3}}^{1/2},
\end{equation}
where $M_{\rm h,12}=M_{\rm h}/10^{12}\,\Msun$ is the halo virial mass. The halo is assumed to follow an NFW profile \citep{NFW}, characterized by a concentration $c$.

We follow \cite{K18} and assume that the radius of the galaxy is proportional to the virial radius, with a constant proportionality factor
\begin{equation}
    R_{\rm d} \approx 0.035 R_{\rm v}.
\end{equation}
This resembles the relation $R_{\rm d}=\lambda R_{\rm v}$, with $\lambda\approx 0.04$ the halo spin parameter \citep{Bullock01}, which follows an assumption of angular momentum conservation during disc formation\footnote{We note that it has been shown in simulations that the size of a given galaxy is not correlated with the spin of its host halo \citep{Jiang19}, because the gas angular momentum is not conserved during its contraction \citep{Danovich15}.} \citep{FallEfstathiou,MMW98}. We assume the galaxy rotation velocity to be proportional to the halo maximum circular velocity, $V_{\rm d} \approx \gamma V_{\rm max}$, which for an NFW halo is given by
\begin{equation}
    V_{\rm max} = 0.465\sqrt{\frac{c}{\ln(1+c)-c/(1+c)}}V_{\rm v}.
\end{equation}
We choose as a fiducial value $c=10$, valid on average for a halo with $M_{\rm h,12}=1$ at $z=1$ \citep{Zhao09}, and note that values of $V_{\rm max}$ for the range $c=1-30$ deviate from each other by less than a factor of two, and this deviation can be absorbed into the parameter $\gamma$. \cite{K18} chose $\gamma=1.4$ quite arbitrarily in order to achieve the correct rotation speed of a Milky Way-like galaxy. It turns out that this choice is in good agreement with cosmological simulations \citep[][figure 7]{DekelRings}, which find $\gamma\approx 1-1.5$ in the relevant halo-mass range for discs (see discussion in \S\ref{sec:Results}). This small scatter introduces small changes to the dynamical time that will not affect our qualitative results. We therefore also adopt $\gamma = 1.4$ as our fiducial value.

\smallskip With the disc radius and rotation velocity in hand, we can compute the dynamical time of the galaxy, given by
\begin{equation}
    \td = \frac{1}{\Omega} = \frac{\Rd}{\Vd} \sim 50-500\, {\rm Myr}.
\end{equation}
\subsection{Disc instability}\label{sec:disc_instability}
The first main assumption in our model is that self-gravitating, rotating discs undergo gravitational instability and self-regulate to marginal instability. The linear stability of a thin, gaseous, self-gravitating, rotating disc, following \cite{Toomre64}, is governed by the Toomre-$Q$ parameter
\begin{equation}\label{eq:Q_parameter}
    \Qg = \frac{\kappa\sigma_{\rm g}}{\pi G\Sigma_{\rm g}},
\end{equation}
where $\Sigma_g=\Mg/\pi\Rd^2$ is the disc gas mass surface density, $\sigma_g$ is the gas radial velocity dispersion, and $\kappa$ is the disc epicyclic frequency. Assuming an isotropic velocity dispersion, $\sigma_g$ is equivalent to the one-dimensional velocity dispersion. For a disc with a rotation curve $v\propto r^{\beta}$, the epicyclic frequency is given by $\kappa = \sqrt{2(1+\beta)}\Omega$. We will focus on discs with flat rotation curves, namely $\beta = 0$. The disc is stable to axisymmetric perturbations when $\Qg>1$. When $\Qg<1$\footnote{If the disc has a finite thickness, the critical Toomre-$Q$ for instability is modified, depending on the anisotropy of the velocity dispersion. For an isotropic velocity dispersion, the critical Toomre-$Q$ for instability is $0.67$ rather than unity \protect\citep{Goldreich65}.}, the disc is unstable to axisymmetric perturbations, and will break up into rings that will eventually fragment into bound clumps.

\smallskip Following \cite{DekelSariCeverino}, we write the Toomre-$Q$ parameter in the form
\begin{equation}
    \Qg = \sqrt{2(1+\beta)}\delta^{-1}\frac{\sigma_{\rm g}}{\Vd},
\end{equation}
where $\delta = \Mg/M_{\rm tot}$, with $\Mg$ being the gas mass within the disc\footnote{The mass that goes into $\delta$ in principle consists of all of the cold mass that participates in the instability, including young stars.}, and $M_{\rm tot}$ is the total mass within a sphere of a radius $\Rd$, including the bulge and dark matter component.

\smallskip The picture of gravitational instability slightly changes when one takes into account both cold material, including cold gas and young stars, together with hotter, older stars as the drivers of instability. When considering such a two component disc, the instability is governed by an effective Toomre-$Q$ parameter, that can be approximated by \citep{Rafikov01,RF13}
\begin{equation}\label{eq:Q_2_comp}
    Q = \left(\Qg^{-1} + \frac{2\sigma_{\rm g}\sigma_{\rm s}}{\sigma_{\rm g}^2+\sigma_{\rm s}^2}Q_{\rm s}^{-1}\right)^{-1},
\end{equation}
where $\Qg$ is given by eq. \ref{eq:Q_parameter}, and $Q_{\rm s}$ is the analogous Toomre-$Q$ parameter for the stellar component\footnote{When performing the linear analysis for a collisionless stellar disc, the $\pi$ in eq. \protect\ref{eq:Q_parameter} is replaced with $3.36$ \protect\citep{Toomre64}, a negligible difference that we ignore here.}. 
The disc of gas and stars is stable whenever $Q\!>\!1$. Following \cite{K18}, we write the two component Toomre-$Q$ parameter as
\begin{equation}
    Q = f_{\rm g,Q}\Qg,
\end{equation}
by defining
\begin{equation}\label{eq:fgq_def}
    f_{\rm g,Q} = \frac{\Mg}{\Mg+2\left[\sigma_{\rm g}^2/(\sigma_{\rm g}^2+\sigma_{\rm s}^2)\right]M_{\rm s}},
\end{equation}
which is an effective gas fraction that coincides with the common gas fraction when $\sigma_{\rm g}\approx \sigma_{\rm s}$. We note that the two component disc can be unstable even when each component is stable had it been a one component disc.

We assume that the discs remain in a state of two-component marginal instability with $Q=1$. The value of $f_{\rm g,Q}$ is near $0.7$ (see \S\ref{sec:redshift_varying_params}) for a gas-rich disc, appropriate for high redshift galaxies \citep{Genzel14}, while it is near a value of $f_{\rm g,Q}=0.5$ in the Solar neighborhood of the Milky Way, as well as star forming galaxies in the xCOLD GASS survey \citep{K18,Xiaoling21}. In \S\ref{sec:redshift_varying_params} we explain how we parameterize the transition from high to low redshifts as a continuous function of redshift.
\subsection{Mass budget}\label{sec:mass_budget}
The gas mass in our model follows a mass conservation equation of the form
\begin{equation}
    \dot{M}_{\rm g} = \dot{M}_{\rm source} - \dot{M}_{\rm sink},
\end{equation}
where $\dot{M}_{\rm source}$ represents the rate at which mass is added to the gaseous disc, while $\dot{M}_{\rm sink}$ represents the rate at which the disc loses gas mass, for example by star formation or by the transport of mass into a bulge. In the following subsubsections, we lay out the different sources and sinks assumed in the model.
\subsubsection{Gas accretion}
Galaxies are subject to accretion of baryons and dark matter from the cosmic web. The halo accretion rate is robustly estimated from Extended Press-Schechter (EPS) theory. We adopt an approximation calibrated to match cosmological simulations, and is given by \citep{ND08}
\begin{equation}\label{eq:halo_accretion_rate}
    \frac{\dot{M}_{\rm h}}{M_{\rm h}} = -aM_{\rm h,12}^b\dot{\omega}.
\end{equation}
Here, $b=0.14$, $a=0.628$ and $\omega$ is the EPS self similar time variable, whose time derivative is given by \citep{ND08}
\begin{equation}
    \dot{\omega}=-0.0476\left(1+z+0.093(1+z)^{-1.22}\right)^{2.5}\,\Gyr^{-1}.
\end{equation}
A good approximation to eq. \ref{eq:halo_accretion_rate} in the EdS cosmological regime, valid for roughly $z\!>\!1$, is given by
\begin{equation}\label{eq:halo_specific_accretion_eds}
    \frac{\dot{M}_{\rm h}}{M_{\rm h}} \approx s(1+z)^{\mu}\ \Gyr^{-1}.
\end{equation}
where $\mu=5/2$. The power $\mu=5/2$ can be simply understood from a scaling argument, based on the Press-Schechter theory \citep{PS}. A key element in the Press-Schechter formalism is the self-invariant time variable, $\omega\propto D(a)^{-1}$, where $D(a)$ is the growth rate of linear perturbations. Self invariance implies that the growth rate of haloes with respect to $\omega$ must be independent of $\omega$, namely ${\rm d}M/{\rm d\omega} = const.$. This can be written as $\dot{M}\propto\dot{\omega}$. In the EdS cosmological regime, $\omega\propto D(a)^{-1}\propto a^{-1}$. With $a\propto t^{2/3}$, this implies that $\dot{M}\propto a^{-5/2} = (1+z)^{5/2}$. At lower redshifts, the power is slightly smaller \citep{ND08}.

\smallskip Equation \ref{eq:halo_specific_accretion_eds} can be integrated analytically, resulting in \citep{Dekel13}
\begin{equation}
    M_{\rm h} = M_{\rm h,z_0}e^{-\tilde{s} (z-z_0)},\ \ \tilde{s}\approx 0.79
\end{equation}
In practice, we numerically integrate eq. \ref{eq:halo_accretion_rate} to generate mass histories of haloes that will host our galaxies.

\smallskip We assume that the amount of accreted baryons in the overall accretion is $f_{\rm b}\dot{M}_{\rm h}$, where $f_{\rm b}\approx 0.17$ is the universal baryonic fraction \citep{Dekel13}. However, not all of the accreted baryons eventually reach the galaxy, as the penetration of cold material is suppressed when the galaxy supports a virial shock, and flows mainly through cold narrow streams that penetrate the hot halo. This virial shock suppression is expected above halo masses $M_{\rm h,12}\!\gtrsim\!1$ \citep{BD03,DB06}. To take this into account, we use the approximation used by \cite{K18}, based on fits to numerical simulations by \cite{FG11}
\begin{equation}\label{eq:gas_accretion_rate}
    \dot{M}_{\rm g,acc} = \epsilon_{\rm in}f_{\rm b}\dot{M}_{\rm h},
\end{equation}
where $\epsilon_{\rm in}$ is a penetration parameter, parameterized as
\begin{equation}\label{eq:eps_in}
    \epsilon_{\rm in} = \min\left(\epsilon_0M_{\rm h,12}^{\alpha_1}(1+z)^{\alpha_2},1\right).
\end{equation}
The best-fit parameters for $z\!\geq\!2$ are $\epsilon_0=0.31,\alpha_1=-0.25,$ and $\alpha_2=0.38$. {In the mass and redshift range of interest for us here, this fit is good to within a factor of two. At $z\leq1$ and $M_{\rm h} \geq 10^{13}\,\Msun$, the fit may overestimate the true penetration found by \cite{FG11} by up to a factor of three, but haloes of such masses are not expected to host discs (see below), and are therefore not relevant for us here.}
\subsubsection{Star formation}\label{sec:sfr}
Motivated by the Kennicutt-Schmidt law \citep{Schmidt59,Kennicutt98}, we assume that the instantaneous star formation rate (SFR) is proportional to the mass in gas,
\begin{equation}\label{eq:sf_law}
    \dot{M}_{\rm SF} = \epsilon_{\rm ff}\frac{\Mg}{t_{\rm ff}},
\end{equation}
where $t_{\rm ff}$ is the free fall time estimated at the midplane of the disc, and $\epsff$ is the efficiency of star formation per free-fall time. The latter has been extensively studied, both observationally and theoretically \citep{KM05,Nalin16}. The values are usually in the range $\epsff\approx 0.01-0.1$, and depend on the relevant scale at which star formation occurs \citep{DekelSariCeverino}, the timescale over which it is measured and the virialization of the molecular clouds \citep{Kim21}. Here, we adopt a median value of $\epsff = 0.015$, similar to other studies \citep{Bouche10, Lilly13, DM14}. We write the freefall time as $t_{\rm ff} = \chi \td$, and we follow \cite{K18} to find $\chi$, assuming a vertical force balance between self gravity, thermal and non-thermal pressure sources. The resulting proportionality is
\begin{equation}\label{eq:ff_to_td}
    \chi = \frac{\pi \Qg}{4}\sqrt{\frac{3f_{g,P}\phi_{\rm mp}}{{2(1+\beta)}}},
\end{equation}
where $f_{g,P}$ is the fraction of midplane pressure which is due to the self gravity of the gas \citep{OS10}, and $\phi_{\rm mp}$ is the excess of pressure due to non thermal and non turbulent sources (e.g. magnetic fields, cosmic rays, etc.). The former is about one half for local galaxies \citep{K18,Xiaoling21}, assuming that the scale height of the gaseous disc is smaller than that of the stars, while it is expected to be larger at high redshifts, where the gas dominates the self gravity of the disc. Similarly to our treatment of $f_{\rm g,Q}$, we adopt a redshift parameterization that will smoothly change the value of $f_{\rm g,P}$ from $0.7$ at high redshifts to $0.5$ at low redshifts, as explained in \S\ref{sec:redshift_varying_params}. The value of $\phi_{\rm mp}$ is in the range $\phi_{\rm mp}=1-2$, and we adopt $\phi_{\rm mp}=1.4$ as in \cite{K18}.

\smallskip By writing the star formation rate as in eq. \ref{eq:sf_law}, we have neglected two corrections. First, is the fact that at low redshifts, star formation typically occurs locally in giant molecular clouds (GMCs), as opposed to high redshifts, where star formation occurs in the giant clumps which are governed by the Toomre instability of the entire disc. The former is called the `GMC regime`, and the latter the `Toomre regime` \citep{KDM12}. This transition occurs roughly when $t_{\rm ff}\epsff^{-1}\!\sim\! 2\, \Gyr$, above which (namely, in the GMC regime at low redshift where the gas depletion time is long) the relation between the local free fall time to the dynamical in eq. \ref{eq:ff_to_td} is no longer valid. Second, we have neglected the fact that at low redshifts, not all of the gas is in a star forming molecular phase, but rather in a warm atomic phase \citep{K12}. As we show below when comparing the results of our model to that of \cite{K18}, who did take these two effects into account, the resulting difference is less than a factor of two.

\subsubsection{Mass transport}\label{sec:mass_transport}
It is well established that gravitationally unstable, turbulent discs exert torques that transport angular momentum outwards, towards larger radii, which by conservation of angular momentum causes mass transport through the disc from the outskirts to the centre. This gradually drains the disc of its gas and partly its stars as well \citep{Gammie,DekelSariCeverino,KB10,Forbes12,Goldbaum15}. There are several ways to estimate the transport rate, corresponding to different mechanisms that drive the transport. Here, in order to allow for uncertainties in estimating the transport rate, we focus on two alternative mechanisms: (i) mass transport due to non axisymmetric torques induced by clump encounters \citep{DekelSariCeverino}; (ii) mass transport due to turbulent viscosity \citep{KB10}.

\smallskip The mass transport due to torques induced by clumps was calculated by \cite{DekelSariCeverino} by estimating the timescale for encounters between clumps and for interactions between clumps and the interclump medium. The transport rate is estimated by
\begin{equation}\label{eq:dsc_transport}
    \dot{M}_{\rm trans,c} = \frac{\Mg}{t_{\rm evac}},
\end{equation}
where $t_{\rm evac}$ is the disc evacuation time, given by
\begin{equation}\label{eq:dsc_timescale}
    t_{\rm evac} = \alpha_c^{-1}\frac{24}{\sqrt{3}\pi^2}(1+\beta)\Qg^2\delta^{-2}\td = \alpha_c^{-1}\frac{12}{\sqrt{3}\pi^2}\Qg^4\lbrac{\frac{\Vd}{\sigma_{\rm g}}}^2 \td.
\end{equation}
The parameter $\alpha_c$ in eq. \ref{eq:dsc_timescale} is the fraction of cold mass in clumps, which will play a key role in regulating the disc in a marginally unstable, turbulent state.  

\smallskip The mass transport in turbulent viscous discs is derived by solving the hydrodynamical equations when taking into account momentum transfer due to turbulent motions and a constant Toomre-$Q$ parameter \citep{KB10}. The resulting equation is an energy conservation equation, which expresses the balance between turbulence dissipation and turbulence driving. We describe in detail below these different mechanisms, and write here the equation with its full generality. \cite{KB10} found that systems reach a steady state, in which the governing equation becomes
\begin{equation}\label{eq:K18_energy}
    -\frac{1-\beta}{2}\Omega\mathcal{T} = \mathcal{L-G},
\end{equation}
where $\mathcal{T}$ is the viscous torque due to turbulent viscosity, $\mathcal{L}$ is the turbulence dissipation rate and $\mathcal{G}$ is the turbulence driving term. The mass transport rate is found by noting that the viscous torque in a steady state is given by $\mathcal{T} = -\dot{M}_{\rm trans,v}\Vd\Rd$. The resulting transport rate is
\begin{equation}\label{eq:kb_transport}
    \dot{M}_{\rm trans,v} = \frac{2}{\Vd^2\left(1-\beta\right)}\left(\mathcal{L-G}\right).
\end{equation}
It is useful to define a dimensionless viscous torque via the equation $\dot{M}_{\rm trans,v} = \tau \Mg/\td$. Equations \ref{eq:K18_energy} and \ref{eq:kb_transport} show that the disc self regulates the turbulent viscous torque in order to sustain energy equilibrium, as discussed in more detail in \S\ref{sec:QE_turb}.
\subsection{Turbulence budget}\label{sec:turb_budget}
\subsubsection{Turbulence dissipation}
Due to its dissipative nature, gas tends to lose energy by shocks and radiation. Turbulence in gas, and in particular supersonic turbulence, as observed at both high and low redshifts \citep{Ianjamasimanana15,Ubler19}, quickly cascades to smaller scales until reaching the viscous scale at which the turbulence energy dissipates. The turbulence dissipation timescale is of the order of the crossing time of the largest driving scale of the turbulence \citep{MacLow98}. To be explained below, we parameterize the dissipation timescale as
\begin{equation}\label{eq:tdiss}
    \tdiss = \frac{\gamma_{\rm diss}}{\sqrt{2(1+\beta)}}\Qg^n\td.
\end{equation}

\smallskip If the driving scale of turbulence is the scale height of the disc, then $n=1$, and $\gamma_{\rm diss}=1/2$ at low redshifts, while $\gamma_{\rm diss}\approx 1$ at high redshifts. To see this, we first adopt the approximation of \cite{Forbes12} for the scale height
\begin{equation}\label{eq:Hg}
    H_{\rm g} = \frac{\sigma_{\rm g}^2}{\pi G\left(\Sigma_{\rm g}+(\sigma_{\rm g}/\sigma_{\rm s})\Sigma_{\rm s}\right)} = \frac{\sigma_{\rm g}^2}{\pi G\Sigma_{\rm g}\phi_Q},
\end{equation}
where $\phi_{\rm Q} = 1 + Q_{\rm g}/Q_{\rm s}$. This parameterization of the gas scale height interpolates between the limit of the scale height being dominated by the self gravity of the gas ($\sigma_{\rm g}/\sigma_{\rm s}\!\ll\!1$) and the limit of the scale height being determined by the gravity of both stars and gas ($\sigma_{\rm g}\approx \sigma_{\rm s}$). At low redshifts, $\phi_Q=2$ \citep{K18,Xiaoling21}, while it should be smaller for high redshift galaxies, as gas dominates the instability. This implies that $\gamma_{\rm diss}=1/2-1$.

\smallskip Alternatively, if the typical clump scale is the relevant scale for the turbulence dissipation time, then $\gamma_{\rm diss}=3\pi/8\sim1$, and $n=-1$ \citep{DekelSariCeverino}. For $Q\!\sim\!1$ this is a factor of two difference with respect to the disc scale height, however for thick discs this can be larger by a factor of a few.

\smallskip In this study, we focus on galaxies with $Q\!\approx\! 1$, and in order to compare with \cite{K18}, we focus on dissipation on the disc scale height, remembering that the two dissipation timescales do not differ by much for $Q=1$. Thus, we adopt $n\!=\!1$, while evolution of $\gamma_{\rm diss}$ with redshift is outlined in \S\ref{sec:redshift_varying_params}. {We note that an equilibrium value of $Q\!\sim\!0.67$ is possible for thick discs \citep{Goldreich65}, appropriate at high redshifts. In \S\ref{sec:xi_a_of_z}, we discuss how a choice of $Q=0.67$ will affect our results.}
\subsubsection{Gravitational driven turbulence}
The timescale for encounters between clumps to generate turbulence in the disc of level $\sigma_g^2$ was computed by \cite{DekelSariCeverino}, by considering the gravitational cross section for two body interactions between bound clumps. The timescale was found to be
\begin{equation}\label{eq:enc_turb}
    t_{\rm enc,turb} = \frac{36}{\sqrt{3}\pi^2}\alpha_c^{-1}Q_g^4t_{\rm d}.
\end{equation}

\smallskip Alternatively, in the picture of gravitational turbulence driving in a turbulent viscous disc, the relevant timescale for transport to generate a turbulence of level $\sigma_{\rm g}^2$ is simply
\begin{equation}
    t_{\rm trans,turb} = 3\frac{\Mg\sigma_{\rm g}^2}{\dot{M}_{\rm trans,v}V_{\rm d}^2(1-\beta)}.
\end{equation}
Once we impose the energy equilibrium, we will be able to find $\dot{M}_{\rm trans,v}$ and estimate the timescale for this type of transport to generate turbulence (see \S\ref{sec:QE_turb}).
\subsubsection{Supernova feedback turbulence}
In order to estimate the amount of turbulence generated by supernovae, we follow several authors \citep{DekelSilk86,Matzner02,FG13,HH16,K18}. During the final phase of the supernova, after the adiabatic phase is over, a thin shell expands while conserving momentum. The shell then fades into with the surrounding medium when its velocity approaches the velocity dispersion of the surrounding medium \citep[for a summary, see][]{Draine11,Dekel19}. If the initial momentum is $p$, then when the shell mixes with the medium, it adds energy of order $\sim\!p\sigma_{\rm g}$ \citep{Matzner02}. For a star formation rate (SFR) of $\dot{M}_{\rm SF}$ and mean momentum injected per stellar mass formed, appropriately averaged over an initial mass function, of $(p/m)_*$ \citep{OstrikerShetty15}, the energy injection rate is
\begin{equation}\label{eq:Esn_turb_dot}
    \dot{E}_{\rm SN,turb} = \dot{M}_{\rm SF}\pms\sigma_{\rm g}.
\end{equation}
From this expression, we can estimate the timescale for supernova feedback to drive turbulence of level $\sigma_{\rm g}^{2}$, plugging the SFR from eq. \ref{eq:sf_law} and $\chi$ from eq. \ref{eq:ff_to_td},
\begin{equation}
    t_{\rm turb,SN} = \frac{3}{2}\epsilon_{\rm ff}^{-1}\chi\pms^{-1}\sigma_{\rm g}\td.
\end{equation}
Simulations of non clustered supernova suggest that ${(p/m)_*\!\approx\!3000\ {\rm km\cdot s^{-1}}}$ \citep{OS10,HH16}, which we adopt as our fiducial value. {The clustering of supernovae may have a significant effect on $(p/m)_*$, though it is fairly uncertain, ranging from a factor of two above our fiducial value to slightly below it \citep{Walch15,Gentry17,Kim17,Gentry19}. A factor of two increase in the adopted $(p/m)_*$ will double the contribution of feedback to the driving of turbulence. This will affect our qualitative results only for discs in halos of mass $\sim\!10^{12}\,\Msun$ at $z\!\sim\!2-3$}.
\subsubsection{Accretion driven turbulence}\label{sec:acc_driv_turb}
The main form in which fresh gas is supplied to massive galaxies at high redshifts, $M_{\rm v}\!>10^{12}\,\Msun$ and $z\!>\!1$, is in narrow cold streams \citep{DB06,Dekel09,Danovich12}. Long lived discs are expected to exist in this mass range at all redshifts \citep{Dekel21}. When the streams hit the disc, some of the kinetic energy they carry is converted into heat in an accretion shock, while the rest remains in kinetic form, eventually turning into turbulence. The fraction of the initial kinetic energy that contributes to turbulence depends strongly on the density contrast between the colliding flows of streams and discs \citep{KH10}.

\smallskip A series of papers studying simulations of idealized, cosmologically motivated streams \citep{Aung19,M20_cooling} as well as analytic work \citep{M18_GC,M20} suggest that the streams should fragment and form dense clumps before reaching the galaxy. Upon impact, such dense clumps have a high efficiency in converting their bulk kinetic energy into turbulent kinetic energy within the disc \citep{KH10}.

\smallskip Streams are expected to be more efficient at driving turbulence than the more spherical accretion typical at low redshifts. Even if the streams were smooth (i.e. not clumpy), they are much denser and concentrated than a spherical accretion, for a given accretion rate. We therefore expect turbulence driven by clumpy accretion to be most relevant at high redshifts.

\smallskip In this paper, we want to determine the mass and redshift range within which this energy source can be important for driving turbulence, and leave the detailed modelling of how clumpy accretion drives turbulence in rotating discs to future work. We therefore, similarly to \cite{Genel12}, parameterize the rate at which streams provide energy to turbulence as some fraction $\xi_a$ of the total kinetic energy carried by the streams, namely
\begin{equation}\label{eq:E_acc}
    \dot{E}_{\rm acc,turb} = \frac{1}{2}\xi_a\dot{M}_{\rm g,acc}\Vd^2,
\end{equation}
where we have approximated the instreaming velocity as the virial velocity, which is comparable to the rotation velocity of the disc (see \S\ref{sec:gal_halo_con}). According to \cite{KH10}, $\xi_a\!\sim\!\Delta$, where $\Delta$ is the density contrast between the accreting material, either smooth stream material or clumps within the streams, and the disc. Using eq. \ref{eq:E_acc} and eq. \ref{eq:Q_parameter} we derive a timescale for accretion to drive turbulence of level $\sigma_{\rm g}^2$
\begin{equation}\label{eq:tacc}
    \tacc = \frac{3\sqrt{2(1+\beta)}}{\xi_a G\dot{M}_{\rm g,acc}Q_g}\sigma_{\rm g}^3\td.
\end{equation}
Below, we explore the effect of three constant values for the conversion efficiency, representing low level ($\xi = 0.3$), moderate level ($\xi_a=0.6$) and maximal ($\xi_a=1$) efficiencies, and we will see that they result in qualitatively different outcomes. {The chosen values for $\xi_a$ are motivated by crude earlier studies of clumpiness in streams \citep{DekelSariCeverino,Aung19}, to be properly evaluated as a function of halo mass and redshift in future work.} We will also explore the effects of a time-varying conversion efficiency parameter on the evolution of the turbulence in discs.
\subsection{Turbulent energy equilibrium}\label{sec:QE_turb}
We now apply the second main assumption of the model, namely that turbulence is in an equilibrium state, in which the dissipation of turbulence is balanced by the three drivers. This condition can be written as
\begin{equation}\label{eq:QE_turb}
    \dot{E}_{\rm turb,trans} + \dot{E}_{\rm turb,SN} + \dot{E}_{\rm turb,acc} = \dot{E}_{\rm diss}.
\end{equation}
Equation \ref{eq:QE_turb} can be written for both pictures of transport, resulting in
\begin{subequations}\label{eq:QE_both}
\begin{align}
    \frac{\sqrt{3}\pi^2}{36}\alpha_c \Qg^{-4} + \frac{2}{3}\epsff\pms\chi^{-1}\sigma_{\rm g}^{-1} & + \frac{1}{3}\xi_a\frac{G\dot{M}_{\rm g,acc}\Qg}{\sqrt{2(1+\beta)}}\sigma_{\rm g}^{-3} \notag \\
    &=\sqrt{2(1+\beta)}\gamma_{\rm diss}^{-1}\Qg^{-n},\label{eq:QE_enc} \\ 
    \tau\frac{1-\beta}{2}\Vd^2 + \pms \epsff\chi^{-1}\sigma_{\rm g}&+\frac{1}{2}\xi_a\frac{\dot{M}_{\rm g,acc}\td}{\Mg}\Vd^2 \notag \\ 
    & = \frac{3}{2}\sqrt{2(1+\beta)}\gamma_{\rm diss}^{-1}\Qg^{-n}\sigma_{\rm g}^2, \label{eq:QE_visc}
\end{align}
\end{subequations}
Eq. \ref{eq:QE_enc} is the energy balance equation when the transport is governed by clump encounters, while eq. \ref{eq:QE_visc} is the energy balance equation when the transport is governed by viscous torques.

\smallskip By imposing the equilibrium turbulent state in equations \ref{eq:QE_enc} and \ref{eq:QE_visc}, the disc self-regulates the gravitational-driven turbulence by adjusting the relevant parameter in each picture. If we assume a fixed $Q=f_{\rm g,Q}\Qg$, and that $\epsff$ and $\left(p/m\right)_*$ are constant in equations \ref{eq:QE_enc} and \ref{eq:QE_visc}, then in the picture of clump encounters, the disc self-regulates $\alpha_c$, the instantaneous fraction of mass in clumps (or equivalently, the number of clumps) in the galaxy, and in the picture of turbulent viscosity, the disc self regulates the viscous torque $\tau$. By isolating $\alpha_c$ from eq. \ref{eq:QE_enc} and $\tau$ from eq. \ref{eq:QE_visc}, we can write
\begin{subequations}\label{eq:SR_both}
\begin{align}
        \alpha_c&=\frac{36\Qg^{4}}{\sqrt{3}\pi^2}F(\sigma_{\rm g}), \label{eq:SR_alpha} \\
        \tau &= \frac{3}{1-\beta}\lbrac{\frac{\sigma_{\rm g}}{\Vd}}^2F(\sigma_{\rm g}), \label{eq:SR_tau}
\end{align}
\end{subequations}
where
\begin{equation}\label{eq:F_of_sigma}
    F(\sigma_{\rm g}) = \sqrt{2(1+\beta)}\gamma_{\rm diss}^{-1}\lbrac{\frac{Q}{f_{\rm g,Q}}}^{-n}\lbrac{1-\frac{\sigma_{\rm SN}}{\sigma_{\rm g}}-\lbrac{\frac{\sigma_{\rm acc}}{\sigma_{\rm g}}}^3}.
\end{equation}

By writing equations \ref{eq:SR_alpha} and \ref{eq:SR_tau} in this form, we have defined two velocities, $\sigma_{\rm SN}$ and $\sigma_{\rm acc}$, which are given by
\begin{equation}\label{eq:sigma_sn}
    \sigma_{\rm SN} = \frac{2}{3\sqrt{2(1+\beta)}}\pms\epsff\chi^{-1}\gamma_{\rm diss}\lbrac{\frac{Q}{f_{\rm g,Q}}}^{n},
\end{equation}
which is $\sigma_{\rm sf}$ as defined in \cite{K18}, and
\begin{equation}\label{eq:sigma_acc}
    \sigma_{\rm acc} = \lbrac{\frac{\xi_aG\dot{M}_{\rm g,acc}Q^{1+n}\gamma_{\rm diss}}{6(1+\beta)f_{\rm g,Q}^{1+n}}}^{1/3}.
\end{equation}
The ratio $\sigma_{\rm SN}/\sigma_{\rm g}$ represents the fraction of turbulent energy that is sustained by supernova feedback, while the ratio $\lbrac{\sigma_{\rm acc}/\sigma_{\rm g}}^3$ represents the fraction of turbulent energy that is sustained by accretion.\footnote{One might expect these fractions to be scale as $\sigma_{\rm g}^2$, however under the assumption of constant $Q$, supernova driving scales as $\sigma_{\rm g}^2$, while accretion driving does not scale with $\sigma_{\rm g}$. Since the turbulent energy scales as $\sigma_{\rm g}^3$ under constant $Q$, we get the linear and cubic scalings as mentioned in the text.} The velocities $\sigma_{\rm SN}$ and $\sigma_{\rm acc}$ depend on the redshift-dependent parameters $f_{\rm g,Q}, f_{\rm g,P}$ and $\gamma_{\rm diss}$. We discuss the adopted redshift parameterization of these parameters in \S\ref{sec:redshift_varying_params}, and quote typical values for $\sigma_{\rm SN}$ and $\sigma_{\rm acc}$ there.

\smallskip Equation \ref{eq:F_of_sigma} suggests that there is a critical gas velocity dispersion, $\sigma_c(M_{\rm h},z)$, below which eqs. \ref{eq:SR_alpha} and \ref{eq:SR_tau} no longer hold. Below this critical velocity dispersion, neither of the sources of mass transport discussed here produce turbulence. This means that at this velocity dispersion, mass transport is no longer needed by the disc to sustain turbulence, and the transport therefore shuts-off. At this point, in order to maintain energy equilibrium, as in eqs. \ref{eq:QE_enc} and \ref{eq:QE_visc}, the disc would have to self regulate a different parameter.  As discussed in \S\ref{sec:shutoff_of_transport}, this can be either $Q$ or $\epsff$. $\sigma_c$ can be calculated analytically, and this is done in Appendix \ref{sec:sigma_c_app}. It is in the range $\sigma_c\approx 20-100\,\kms$, depending on $\xi_a$, mass and redshift.

\begin{table}
 \caption{Summary of typical values of model parameters.}
 \label{tab:params}
 \begin{threeparttable}
 \begin{tabular*}{\columnwidth}{@{}l@{\hspace*{10pt}}l@{\hspace*{3pt}}l@{\hspace*{5pt}}l@{}}
  \hline
  Parameter & Value & Definition & Meaning\\
  \hline
  $c$ & $10$ & \S\protect\ref{sec:gal_halo_con} & Halo concentration parameter \\[2pt] 
  $\gamma$ & $1.4$ & \S\protect\ref{sec:gal_halo_con} & $V_{\rm d}$ in units of $V_{\rm max}$\\[2pt]
  $\beta$ & $0$ & \S\protect\ref{sec:disc_instability} & Rotation curve log slope\\[2pt]
  $Q$ & $1$ & Eq. \protect\ref{eq:Q_2_comp} & Two component Toomre-$Q$ \\[2pt]
  $f_{\rm g,Q}$ & $0.5-0.7$ & Eq. \protect\ref{eq:fgq_def} & Effective gas fraction for Toomre-$Q$\\[2pt]
  $\epsff$ & $0.015$ & Eq. \protect\ref{eq:sf_law} & Star formation efficiency \\[2pt]
  $\chi$ & $1-1.5$ & Eq. \protect\ref{eq:ff_to_td} & Free-fall to dynamical time ratio \\[2pt]
  $f_{\rm g,P}$ & $0.5-0.7$ & \S\protect\ref{sec:sfr} & Fraction of self gravity due to gas\\[2pt]
  $\phi_{\rm mp}$ & $1.4$ & \S\protect\ref{sec:sfr} & Total-to-turbulent pressure ratio\\[2pt]
  $\alpha_c$\tnote{a} & $0-1$  & \S\protect\ref{sec:mass_transport} & Mass fraction in clumps\\[2pt]
  $\tau$\tnote{a} & $0 - 10^{-2}$  & \S\protect\ref{sec:mass_transport} & Dimensionless viscous torque\\[2pt]
  $\gamma_{\rm diss}$ & $0.5$ & Eq. \protect\ref{eq:tdiss} & Turbulence dissipation parameter \\[2pt]
  $\phi_Q$ & $1.5-2$ & Eq. \protect\ref{eq:Hg} & Stellar to gaseous Toomre-$Q$ ratio\\[2pt]
  $\pms$ & $3000\,\kms$ & Eq. \protect\ref{eq:Esn_turb_dot} & SN momentum per stellar mass\\[2pt]
  $\xi_a$ & $0.3-1$ & Eq. \protect\ref{eq:E_acc} & Accretion-driven turbulence efficiency\\[2pt]
  $\sigma_{\rm SN}$ & $11-14\,\kms$ & Eq. \protect\ref{eq:sigma_sn} & Dispersion sustained by SN\\[2pt]
  $\sigma_{\rm acc}$ & $8-54\,\kms$ & Eq. \protect\ref{eq:sigma_acc} & Dispersion sustained by accretion\\[2pt]
  \hline
 \end{tabular*}
 \begin{tablenotes}
 \item[a] The values of $\alpha_c$ and $\tau$ depend on the gas velocity dispersion, halo mass, redshift and all of the model parameters, see \S\ref{sec:QE_turb}.
 \end{tablenotes}
 \end{threeparttable}
\end{table}
Using the self-regulated parameters, we can write expressions for the mass transport rates in the steady turbulent state. In the picture of clump encounters, the mass transport rate is
\begin{equation}\label{eq:dsc_transport_SR}
    \dot{M}_{\rm trans,c} = 3F(\sigma_{\rm g})\lbrac{\frac{\sigma_{\rm g}}{\Vd}}^2\frac{\Mg}{\td},
\end{equation}
and in the picture of turbulence viscosity,
\begin{equation}\label{eq:kb_transport_SR}
    \dot{M}_{\rm trans,v} = \frac{3}{1-\beta}F(\sigma_{\rm g})\lbrac{\frac{\sigma_{\rm g}}{\Vd}}^2\frac{\Mg}{\td}.
\end{equation}
We can see that up to a factor that depends on the rotation curve, which can introduce a factor of at most two, the resulting mass transport rates are identical. Using this surprising result, in the remainder of the paper we assume a flat rotation curve, $\beta = 0$, ignore the different origins of mass transport, and define
\begin{equation}\label{eq:mass_transport}
    \dot{M}_{\rm trans} = 3F(\sigma_{\rm g})\lbrac{\frac{\sigma_{\rm g}}{\Vd}}^2\frac{\Mg}{\td}
\end{equation}
as the sink term in eq. \ref{eq:mass_conservation}, which is related to the mass transport within the disc. While the expressions for the transport rates are the same, one must remember that their physical origin is different and that the physical meaning of the corresponding self-regulated parameter, $\alpha_c$ and $\tau$, is different in each picture.


\subsection{Redshift evolution of $f_{\rm g,Q}, f_{\rm g,P}$ and $\gamma_{\rm diss}$}\label{sec:redshift_varying_params}
The parameters $f_{\rm g,Q}, f_{\rm g,P}$ and $\gamma_{\rm diss}$ may vary with time. At high redshifts, discs are gas rich and the dynamics is more influenced by the gaseous component. At low redshifts, the discs are gas-poor, in which both the cold gas and the hotter stars contribute the dynamics. The value of $\phi_Q$, which sets the value of $\gamma_{\rm diss}$ (see eq. \ref{eq:Hg}), also varies with time, as the height of the gaseous disc is thicker at high redshifts than low redshifts.

According to \cite{Genzel15}, for star-forming galaxies on the Main Sequence in the redshift range $z=0-4$, the typical gas fraction varies as 
\begin{equation}\label{eq:f_g}
    f_g(z) \approx \frac{0.06\cdot (1+z)^{2.7}}{1+0.06\cdot (1+z)^{2.7}}.
\end{equation}
In order to inspect the qualitative effects of the variation in time of the aforementioned parameters, we make the crude assumption that $f_{\rm g,Q}, f_{\rm g,P}$ and $\phi_Q$ evolve as a function of the gas fraction,
\begin{equation}
    f_{g,Q} = f_{g,P} = \phi_Q^{-1} = a\cdot f_g(z) + b,
\end{equation}
where $a=0.39$ and $b=0.38$ are chosen so that the values will vary between $0.7$ at $z=4$ to $0.5$ at $z=0$. These values are chosen based on the gas fraction of observed galaxies \citep{Genzel15,Tacconi18} and velocity dispersions of gas and stars from simulations \citep{Pillepich19}. We defer to future work a more detailed study using the redshift dependence as derived from simulations.

\smallskip Under this parameterization, $\sigma_{\rm SN}$ evolves from ${\sigma_{\rm SN}\!\approx\! 13\,\kms}$ at $z\!=\!2$ to $\sigma_{\rm SN}\!\approx\! 11\,\kms$ at $z\!=\!0$. For $\xi_a = 0.6$, $\sigma_{\rm acc}\!\approx\! 20, 36, 61\,\kms$ for galaxies in haloes of mass $\log\lbrac{M_{\rm h}/\Msun} = 11.5, 12, 13$ at $z=2$, and $\sigma_{\rm acc}\approx 8, 12, 24\,\kms$ for galaxies in haloes of the same masses at $z=0$.
\section{Results}\label{sec:Results}
Equipped with all of the relevant mass sources and sinks, we can write a conservation equation for the gas mass in the disc
\begin{equation}\label{eq:mass_conservation}
    \dot{M}_{\rm g} = \dot{M}_{\rm g,acc} - \dot{M}_{\rm trans} - \dot{M}_{\rm SF}.
\end{equation}
We integrate equation \ref{eq:mass_conservation} numerically\footnote{We use the ${\tt solve\_ivp}$ method in the ${\tt SciPy}$ package \citep{scipy}, using the BDF method.}, using eqs. \ref{eq:gas_accretion_rate}, \ref{eq:sf_law} and \ref{eq:mass_transport} for the relevant quantities. The mass transported down the potential well ends up in a bulge, which is assumed to be a distinct component of the galaxy, and we neglect its dynamical implications (see \S\ref{sec:discussion} for discussion).

\smallskip Since equation \ref{eq:mass_conservation} is driven by the external accretion, which evolves in time slower than other relevant timescales, {the solution will eventually converge to a unique solution that depends only on the model parameters and the halo mass at redshift zero (due to the model cosmological accretion rate), and not on the particular choice of initial conditions}. This characteristic of the bathtub model has already been investigated in other bathtub studies \citep{Bouche10,Lilly13,Genel12,DM14,Cacciato12}. 

\smallskip Once we obtain the evolution of the gas mass, {we can use the fact that $\sigma_g\propto \Qg\Mg$, along with the marginal Toomre instability condition, to solve for $\sigma_g(t)$}. This will allow us to closely inspect how the turbulence is supported against dissipation. The model parameters and their corresponding fiducial values are summarized in Table \ref{tab:params}.

\begin{figure}
    \centering
    \includegraphics[width=\columnwidth]{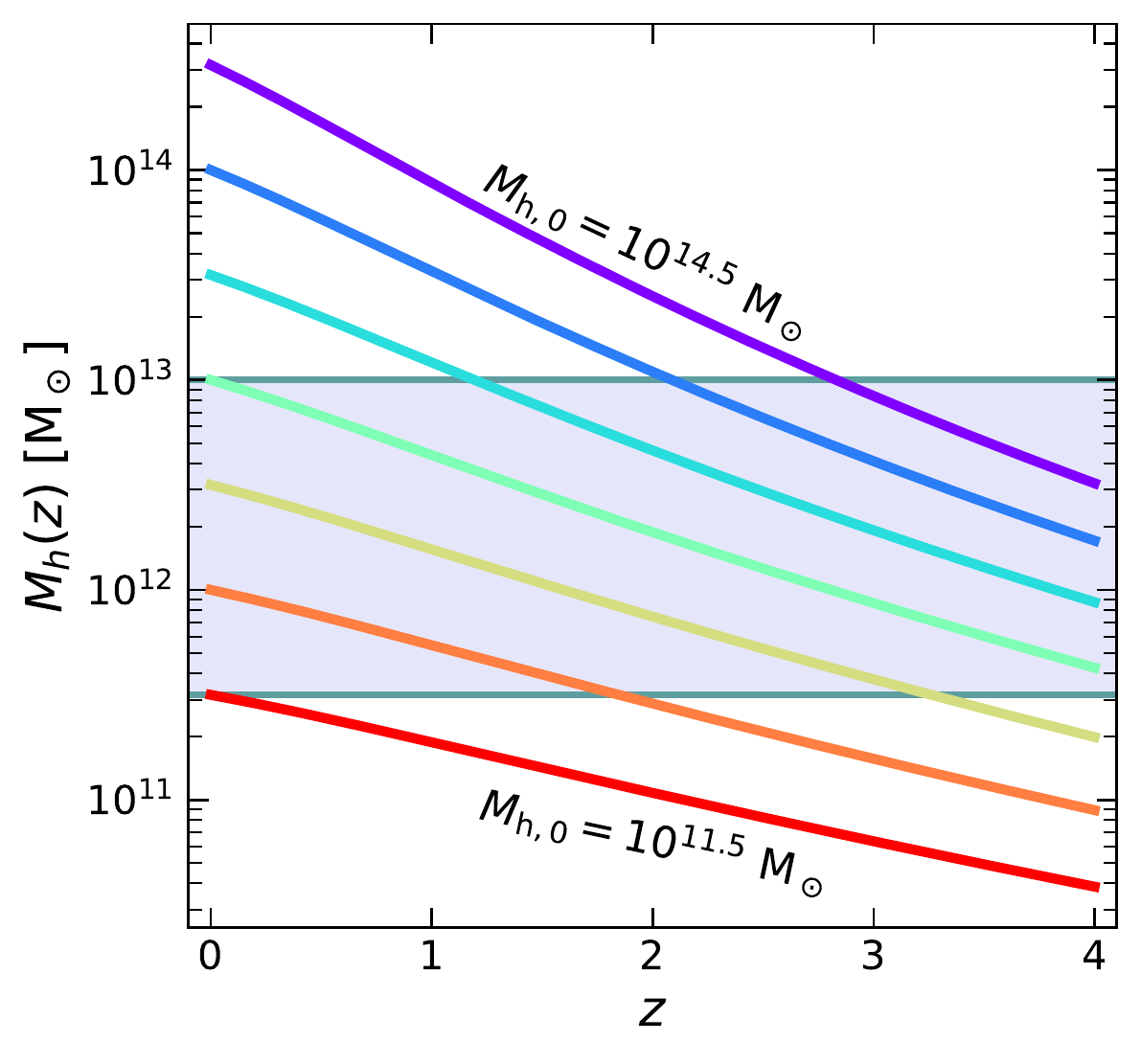}
    \caption{The evolution of the mass of halos hosting discs, as a result of integrating the average accretion rate (eq. \protect\ref{eq:halo_accretion_rate}). Shown are the evolution tracks for halos with mass at redshift $z=0$ between $10^{11.5}\,\Msun$ and $10^{14.5}\,\Msun$, in increments of $0.5\,{\rm dex}$. The shaded band indicates the time the galaxy in each halo is expected to be a disc (see text).}
    \label{fig:halo_mass_evolution}
\end{figure}

\smallskip We follow the evolution of galaxies inside haloes throughout their lifetimes. The evolution of the discs' host halo masses are shown in figure \ref{fig:halo_mass_evolution}. However, the galaxies are not expected to be discs at all times. \cite{DekelDiscs} found that discs tend not to survive for more than an orbital time if they reside in haloes of mass $M_{\rm h}\!<\!10^{11.5}\,\Msun$, roughly independent of redshift. Also, central galaxies inside clusters, i.e. in halos with $\log{M_{\rm h}}\!\gtrsim13$, are expected to be ellipticals rather than discs \citep{Dressler80}. Thus, when presenting our results, we highlight the regions of the parameter space in which we expect to have the galaxies containing extended disks, namely the mass range ${11.5\!\leq\!\log{M_{\rm h}(z)}\!\leq\!13}$ at the given redshift, where our model is applicable. Due to the nature of bathtub models, the assumption that our model is applicable only from a certain point in time does not introduce a significant error, as the solution is not very sensitive to the initial conditions. We find that it takes between one to two disc orbital times, for all initial redshifts, for the solutions to converge to a unique solution, regardless of the initial conditions. We analyze the convergence in Appendix \ref{sec:QSS_app}.

Throughout this section, unless stated otherwise, the quantity $M_{\rm h,0}$ refers to the galaxy's host halo mass at $z\!=\!0$. This mass uniquely determines the entire evolution of the halo mass, assuming the average accretion history given by eq. \ref{eq:halo_accretion_rate}.
\begin{figure}
    \centering
    \includegraphics[width=\columnwidth]{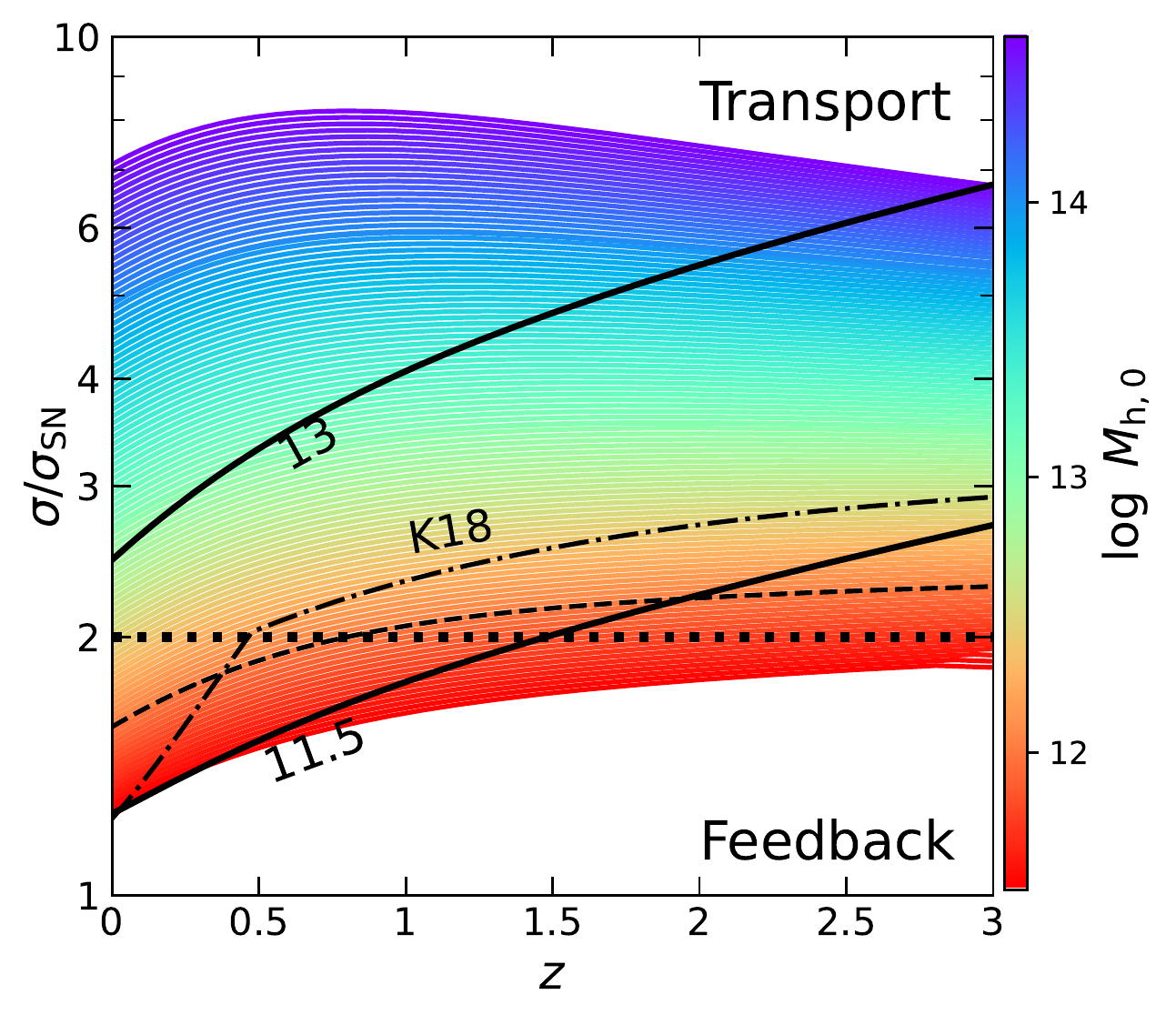}
    \caption{The evolution of gas-disc velocity dispersion, with respect to the contribution of feedback, as predicted by our model, when $\xi_a = 0$, i.e., when accretion is assumed to provide no energy to the turbulence. Here we consider feedback and transport as the drivers of turbulence, and turn off the contribution of accretion. The figure shows the evolution for different galaxies, distinguished by the host halo mass at $z=0$, indicated by the different colors. The dashed line is a sample evolution of a galaxy that resides in a halo with $\log{M_{\rm h,0}}\!=\!12$ today. The dashed dotted line shows the evolution of a galaxy inside a halo of mass $\log{M_{\rm h,0}}=12$ from \protect\cite{K18}. The thick solid black lines bound the region in which our model predicts the galaxies to be discs, namely when the instantaneous halo mass is in the range $11.5\!<\!\log{M_{\rm h}}\!<\!13$. Note that this region is applicable only to the results of our model and not to that of \protect\cite{K18}. Above the thick dotted line, the turbulence in galaxies is mainly supported by transport rather than feedback. We learn that turbulence in all discs is sustained mainly by transport at $z\!\gtrsim1.5$. At lower redshits, $z\!\lesssim\!1$ turbulence in discs that reside in haloes with $\log{M_{\rm h,0}}\!\lesssim\!{12}$ becomes mainly supported by feedback.}
    \label{fig:feedback_transport}
\end{figure}

\subsection{Feedback and transport only}\label{sec:feedback_and_transport}
First, we examine the main driver of turbulence in discs with no accretion, namely $\xi_a=0$, which means $\sigma_{\rm acc}=0$. In this case, the function $F(\sigma_{\rm g})$ that goes into eq. \ref{eq:mass_conservation} is
\begin{equation}\label{eq:trans_no_acc}
    F(\sigma_{\rm g}) \propto 1-\frac{\sigma_{\rm SN}}{\sigma_{\rm g}}.
\end{equation}
Figure \ref{fig:feedback_transport} shows the evolution of the velocity dispersion in the disc, normalized by $\sigma_{\rm SN}$. As discussed in \S\ref{sec:QE_turb}, the ratio $f_{\rm turb,SN}=\sigma_{\rm SN}/\sigma_{\rm g}$ is the fraction of turbulence dissipation that is supported by supernova feedback. So, when $f_{\rm turb,SN} > 1/2$, most of the turbulence is supported by supernova feedback, while when $f_{\rm turb,SN}<1/2$, most of the turbulence is supported by transport.

\smallskip The region of the parameter space in which galaxies are expected to be discs is bounded by the two solid curves. We can see from figure \ref{fig:feedback_transport} that at high redshifts, roughly $z\!>\!1.5$, the turbulence in all discs is expected to be supported by transport. At lower redshifts, turbulence in discs that reside in haloes of mass $\log{M_{\rm h,0}}\!\lesssim\!12$ becomes mainly supported by supernova feedback.

\smallskip Figure \ref{fig:feedback_transport} also shows the evolution of the velocity dispersion from \cite{K18} for a halo with a mass of $10^{12}\ {\rm M_\odot}$ at redshift zero. Note that this result assumes $f_{g,Q}=f_{g,P}=\phi_Q^{-1}=1/2$ with no redshift evolution. In that case $\sigma_{\rm SN} = 11\,\kms$ at all redshifts\footnote{In \protect\cite{K18}, $\sigma_{\rm SN}$ has a weak redshift dependence at $z\!\lesssim\!0.5$, that comes from the evolution of the fraction of gas that is in a molecular phase rather than an atomic phase.}. The results of \cite{K18} suggest that discs that reside in haloes of mass $\log{M_{\rm h,0}} = 12$ have turbulence that is primarily supported by transport at all redshifts until $z\!\sim\!0.5$. However, such galaxies are not expected to be discs until $z\!\sim\!2$, and our model predicts that these galaxies, when they become discs, have turbulence that is supported by supernova feedback and transport with comparable power, rather than predominantly supported by transport. We note that changes to the redshift parameterization of $f_{\rm g,Q}, f_{\rm g,P}$ and $\phi_Q$ as discussed in \S\ref{sec:redshift_varying_params} can change the conclusions only quantitatively, and we still expect that qualitatively, discs with $\log{M_{\rm h,0}}\lesssim12$ will have turbulence that is primarily supported by supernova feedback.

\smallskip We note two differences between our approach to the solution and the one used by \cite{K18}, beyond the parameterization of $f_{\rm g,Q}, f_{\rm g,P}$ and $\phi_Q$. First, \cite{K18} produced the cosmological evolution of the velocity dispersion by solving the equation $\dot{M}_{\rm g,acc} = \dot{M}_{\rm trans}$ for $\sigma_{\rm g}$. The left hand side introduces the cosmological dependence. The reasoning behind this approach is the fact that the bathtub models for galaxy evolution, driven by an external accretion, reach a steady state solution that is dictated by this external source, as discussed above. This steady state solution can be approximated analytically by imposing $\dot{M}_{\rm g}=0$ in equation \ref{eq:mass_conservation}, however \cite{K18} neglected the SFR as a contribution to the sink term. 

\smallskip Second, we have not included either the GMC regime of star formation nor the fact that, at low redshifts, the fraction of gas in the star forming phase is lower. Both of these effects change the mass transport rate, and hence affect the evolution of $\sigma_{\rm g}$. These two corrections introduce a small difference of less than a factor of two, and do not introduce any qualitative difference.

\subsection{Adding accretion}
We now investigate the dominant source of turbulence when adding accretion as a third driver. We will first assume a constant value for $\xi_a$, and investigate the evolution assuming three values for the efficiency of converting accreted kinetic energy into turbulence (as defined in \S\ref{sec:acc_driv_turb}): low efficiency, $\xi_a=0.3$, namely a third of the kinetic energy provided by streams is converted into turbulence, moderate, $\xi_a = 0.6$, and maximal efficiency, $\xi_a=1$. In \ref{sec:xi_a_of_z}, we will investigate very crudely the evolution assuming a redshift-dependent conversion efficiency.
\subsubsection{Accretion and transport only}\label{sec:accretion_transport_only}
First, we consider models with only transport and accretion, namely no feedback ($\sigma_{\rm SN}=0$). The function $F(\sigma_{\rm g})$ that goes into the mass transport rate in eq. \ref{eq:mass_conservation} is in this case
\begin{equation}\label{eq:trans_no_fdbk}
   F(\sigma_{\rm g}) \propto 1-\lbrac{\frac{\sigma_{\rm acc}}{\sigma_{\rm g}}}^3.
\end{equation}
Analogously to the previous section, we will focus on the ratio $f_{\rm turb,acc} = \lbrac{\sigma_{\rm acc}/\sigma_{\rm g}}^3$, which is the fraction of turbulence dissipation that is balanced by accretion driven turbulence. 

\begin{figure}
    \centering
    \includegraphics[width=\columnwidth]{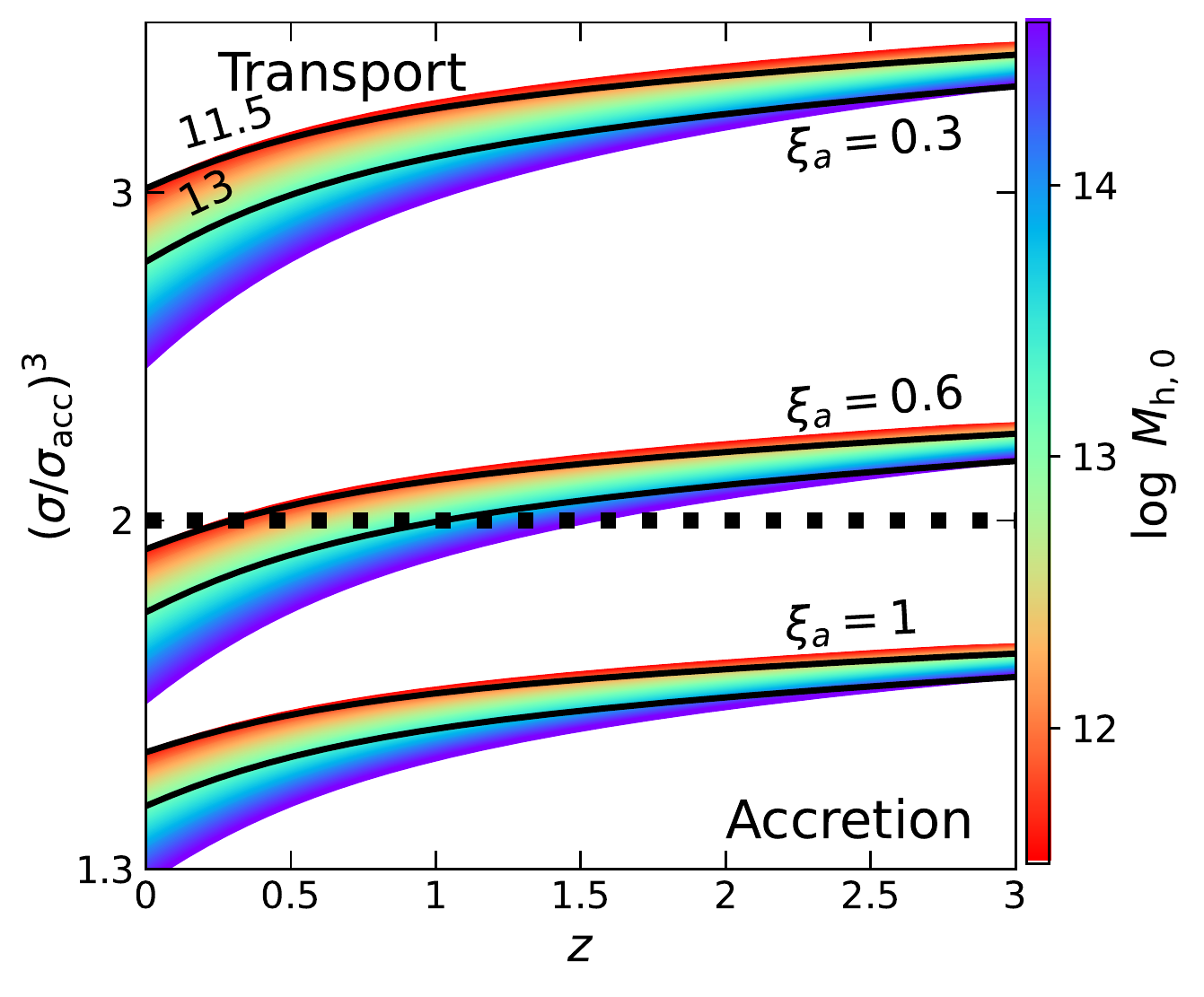}
    \caption{The evolution of $\lbrac{\sigma_{\rm g}/\sigma_{\rm acc}}^3$, which corresponds to the fraction of turbulent dissipation balanced by accretion driven turbulence (see \S\protect\ref{sec:QE_turb}). This figure is analogous to figure \protect\ref{fig:feedback_transport}. When $\lbrac{\sigma_{\rm g}/\sigma_{\rm acc}}^3>2$, mass transport is the primary driver of turbulence. Shown are three sets of evolutionary curves, resulting from a different accretion conversion efficiency: low ($\xi_a=0.3$; top), moderate ($\xi_a=0.6$; middle) and maximal ($\xi_a=1$; bottom). The solid black curves bound the regions in each set where the galaxies are expected to be discs. We learn that for a low accretion conversion efficiency, turbulence in all discs at all times is supported mainly by transport. For a moderate efficiency, turbulence in discs is supported by accretion and transport with comparable power. For a maximal efficiency, all discs are supported mainly by accretion, at all times.}
    \label{fig:accretion_transport}
\end{figure}

\smallskip Figure \ref{fig:accretion_transport} shows the evolution of the cube of $\sigma_{\rm g}$ normalized by $\sigma_{\rm acc}$, for three levels of accretion efficiency, namely $\xi_a=0.3,0.6$ and $\xi_a=1$. Unlike the case with feedback and transport only (\S\ref{sec:feedback_and_transport}), now the normalization, namely $\sigma_{\rm acc}$, is mass dependent. We see from figure \ref{fig:accretion_transport} a qualitative difference between the three levels of conversion efficiency. For $\xi_a=0.3$, we see that all discs have ${\lbrac{\sigma_{\rm g}/\sigma_{\rm acc}}^3\!\sim\!3}$, namely more than $60\%$ of their turbulence is supported by transport, for all masses and redshifts. For $\xi_a=0.6$, all discs have $\lbrac{\sigma_{\rm g}/\sigma_{\rm acc}}^3\!\sim\!2$, namely comparable share between transport and accretion. Lastly, we see from figure \ref{fig:accretion_transport} that for ${\xi_a=1}$, namely maximal conversion efficiency of accretion kinetic energy to turbulence, accretion is the main driver of turbulence in discs of all masses at all times.

\smallskip We notice weak variations of $\lbrac{\sigma_{\rm g}/\sigma_{\rm acc}}^3$ with halo mass, with the ratio decreasing with increasing $\log{M_{\rm h,0}}$. This results from the dependence of the gas accretion rate on halo mass. In our model, this dependency comes from the cosmological accretion rate, as defined in eq. \ref{eq:halo_accretion_rate}, and the penetration parameter, eq. \ref{eq:eps_in}. A constant penetration, as well as a weaker mass dependence of the accretion rate, might further weaken and even invert this gradient. For most reasonable choices, the dependence of $f_{\rm turb,acc}$ on $M_{\rm h}$ is very weak, and the parameterizations of $\dot{M}_{\rm h}$ as a function of $M_{\rm h}$ will not qualitatively affect our conclusions.

\smallskip The results of this section indicate that turbulence driven by accretion can have a significant role in balancing turbulence dissipation. We do note, however, that while different conversion efficiencies $\xi_a$ give qualitatively different results, the fraction of turbulence dissipation that is balanced by accretion is always in the range of $0.4-0.7$, namely its contribution to the total turbulent energy budget is comparable to that of transport. This is unlike the model with feedback and transport, ignoring accretion (\S\ref{sec:feedback_and_transport}), where massive haloes were predominantly supported by transport.
\begin{figure*}
    \includegraphics[width=1.5\columnwidth]{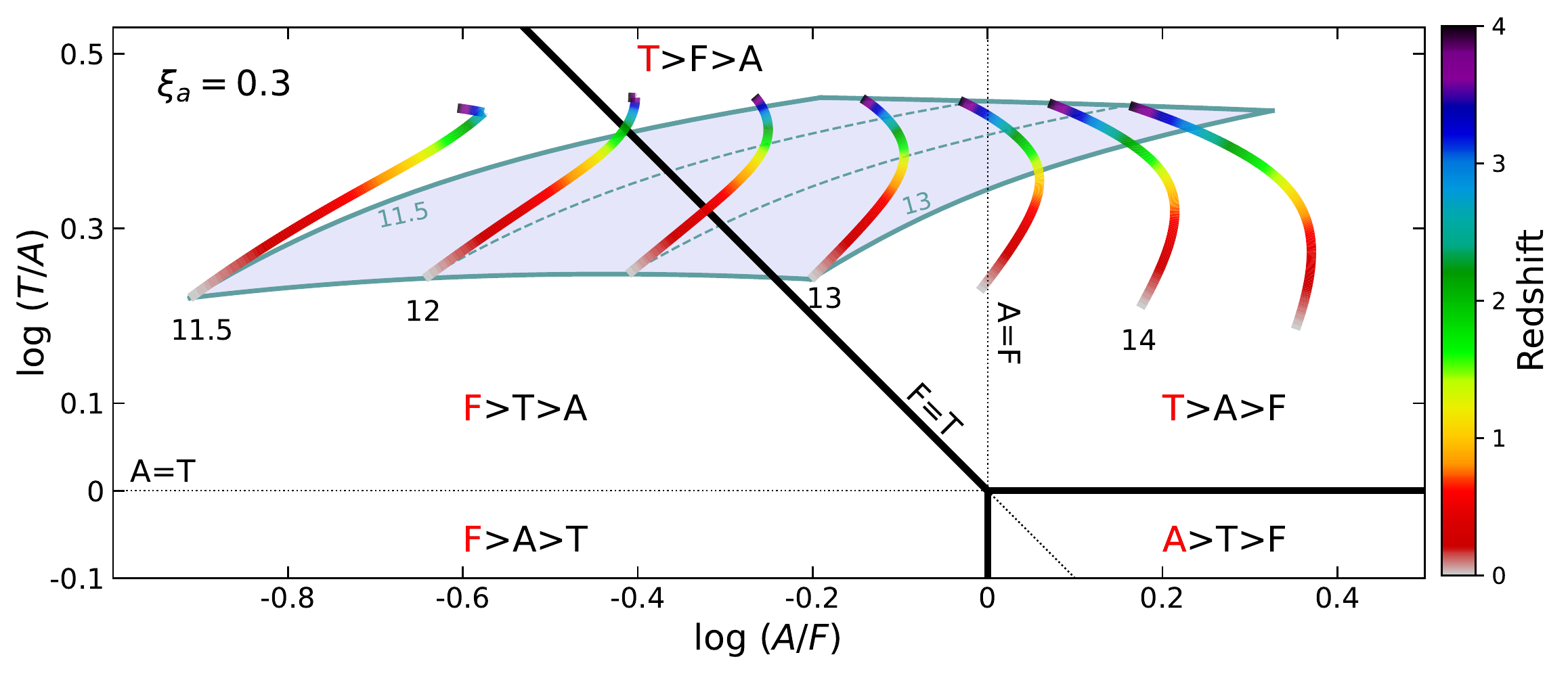} \\
    \includegraphics[width=1.52\columnwidth]{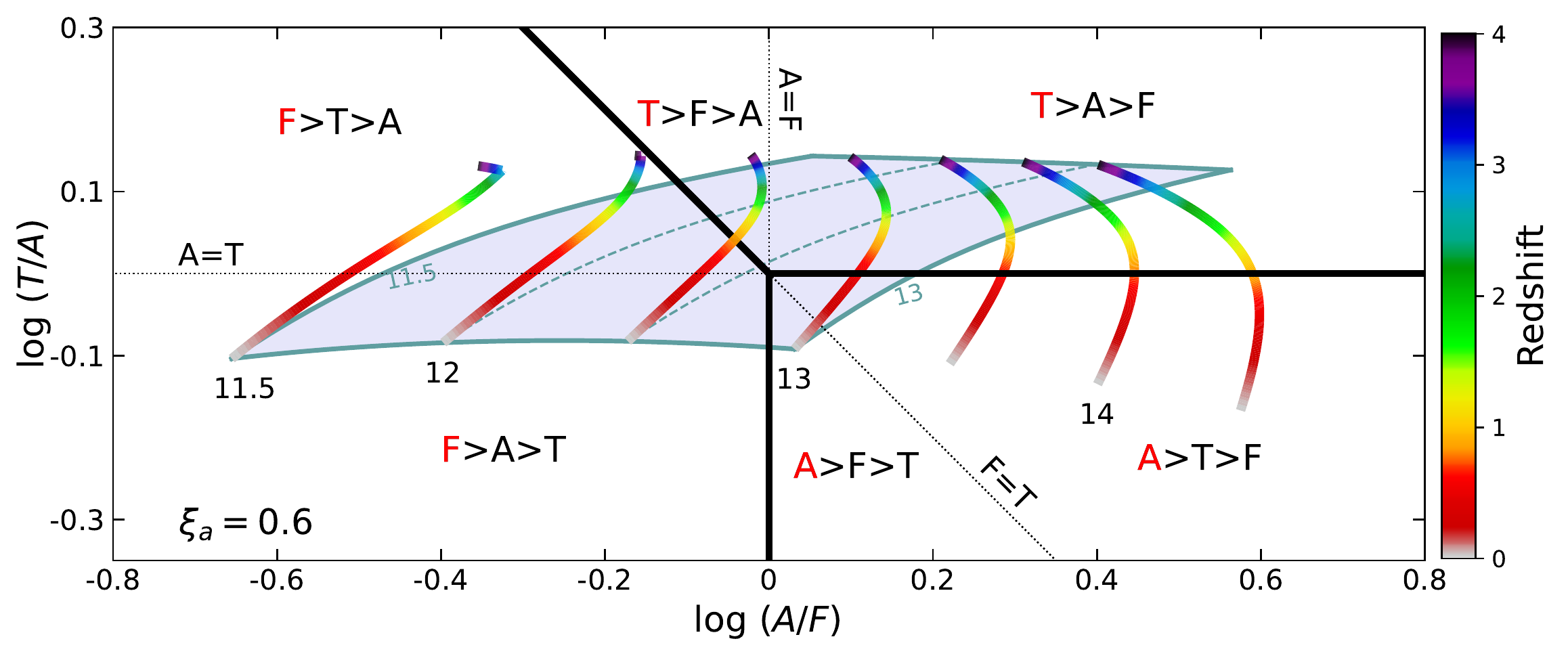}\hspace*{0.2cm} \\
    \includegraphics[width=1.5\columnwidth]{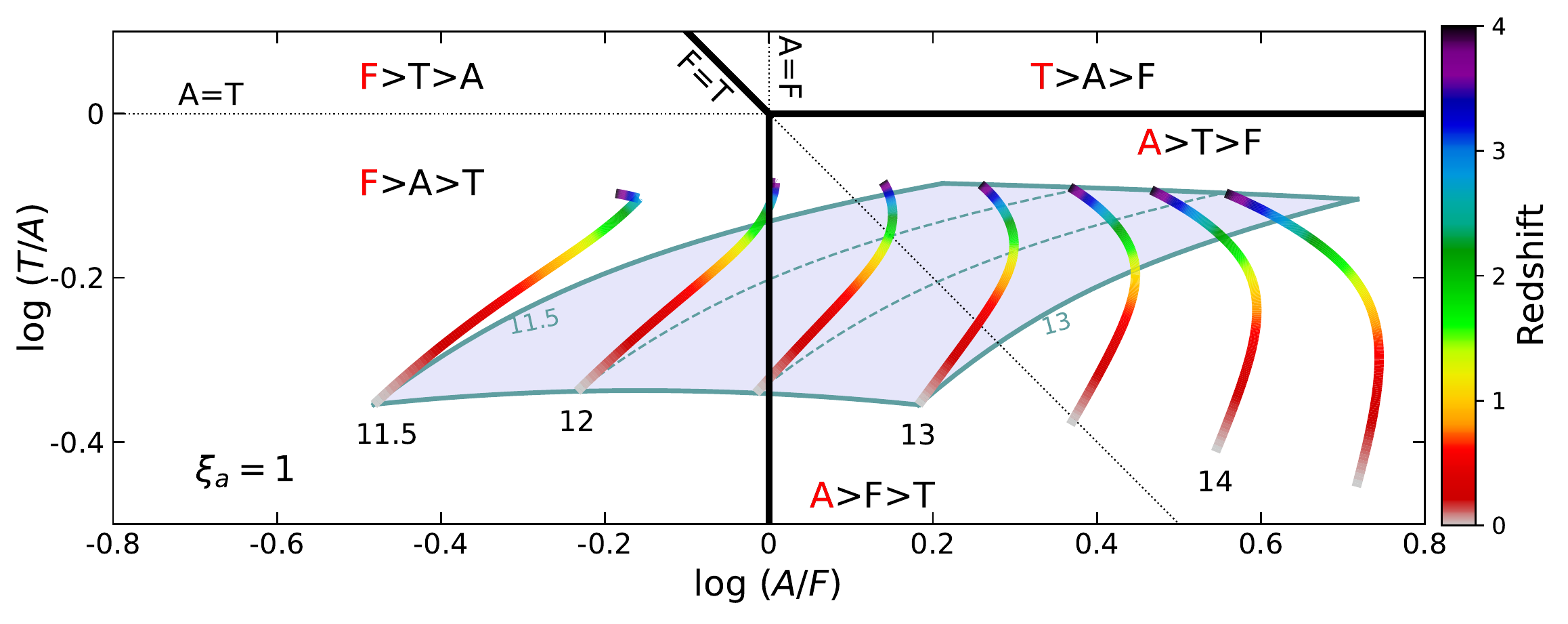}
    \caption{The relative roles of the three processes that drive disc turbulence as a function of halo mass and redshift. $A$, $F$ and $T$ refer to Accretion, Feedback and Transport. The axes refer to the ratios of energy injection rates, $T/A$ and $A/F$. Shown are evolutionary tracks of discs that reside in haloes of different masses at $z=0$, $\log{M_{\rm h,0}}\!=\!11.5-14.5$ in increments of $0.5\,{\rm dex}$. The color indicates the redshift along the track at the particular position in the plane. Thick solid black lines divide the plane to three regions with different dominant drivers, marked by a red letter. The thin lines separate regions with a given primary driver based on their secondary driver. The grey region bounded by the turquoise frame marks the mass range where the galaxies are expected to be discs at the given redshift, $\log{M_{\rm h}(z)} = 11.5-13$. The dashed lines are lines of constant halo mass of $10^{12}\,\Msun$ and $10^{12.5}\,\Msun$. The different panels show the evolution of discs for different values of the accretion conversion efficiency - $\xi_a=0.3$ (top), $\xi_a=0.6$ (middle) and $\xi_a=1$ (bottom). We can see that discs in haloes that have $\log{M_{\rm h,0}}\lesssim12$ today are mainly supported by feedback throughout their lifetimes, regardless of the accretion conversion efficiency. More massive discs are primarily supported by transport for a low conversion efficiency, while for a moderate conversion efficiency we can see a transition from transport dominated turbulence to accretion dominated turbulence near $z\!\sim\!2$.  For a high accretion conversion efficiency, massive discs are always accretion dominated.}
    \label{fig:accretion_tracks}
\end{figure*}
\subsubsection{Accretion, feedback \& transport - constant $\xi_a$} 
Finally, we consider the case where all three drivers of turbulence are active. First, we consider three constant efficiency parameters for converting accretion kinetic energy into turbulence. In this case, the function $F(\sigma_{\rm g})$ that goes into eq. \ref{eq:mass_conservation} is given by eq. \ref{eq:F_of_sigma}. In order to find out which is the primary driver of turbulence as a function of halo mass and redshift, we compare the energy injection rates of each driver, as derived in \S\ref{sec:turb_budget}.

\smallskip Figure \ref{fig:accretion_tracks} presents our main result. It shows evolutionary tracks of galaxies that reside in haloes of different masses at $z=0$, from $\log{M_{\rm h,0}}=11.5$ to $\log{M_{\rm h,0}}=14.5$, in increments of $0.5\,{\rm dex}$. The tracks are plotted in the plane of $\log\lbrac{T/A}$ vs. $\log\lbrac{A/F}$, where $T,A$ and $F$ are the turbulent energy injection rates of Transport, Accretion and Feedback, respectively. The plane has been divided into several regions of two types - the thick lines separate regions of different primary drivers, while the thin lines separate regions of a given primary driver which have different secondary drivers. The shaded region marks the mass range where galaxies are expected to be discs at each redshift, as explained above. We learn from figure \ref{fig:accretion_tracks} the following:

\begin{itemize}
\item For a low accretion conversion efficiency, $\xi_a\!=\!0.3$ (top panel), discs in haloes with $\log{M_{\rm h,0}}\!\lesssim\! 12$ today, are predominantly supported by feedback at all times, while haloes with $\log{M_{\rm h,0}}\gtrsim13$ are mainly supported by transport at all times. Turbulence support in intermediate mass galaxies transitions from being dominated by transport at high redshifts to feedback at low redshifts. This result is qualitatively similar to the result of the model with feedback and transport only (\S\ref{sec:feedback_and_transport}). Furthermore, for $\log{M_{\rm h,0}}\!<\!12$, the secondary driver is transport, with the accretion less important. For massive discs, with $\log{M_{\rm h,0}}\!\gtrsim\!13.5$, corresponding to ${M_{\rm h}(z=2)\approx 10^{12.7}\,\Msun}$, accretion becomes the secondary driver, with a higher energy injection rate than feedback.

\item For a moderate accretion conversion efficiency, $\xi_a\!=\!0.6$ (middle panel), low mass discs, residing in haloes with $\log{M_{\rm h,0}} \lesssim 12$, have their turbulence primarily driven by feedback throughout their lifetimes, similar to the $\xi_a=0.3$ case and the feedback and transport only model. The secondary driver in these discs is transport until $z\!\sim\!0.5$, with accretion becoming the secondary driver at later times. However, this difference is small (see below).

\smallskip Discs that reside in haloes with $\log{M_{\rm h,0}}\!\gtrsim\!12$ at $z\!=\!0$, have their turbulence driven primarily by transport at high redshifts. However, we note that the transport energy injection rate and accretion energy injection rate for all of the galaxies with $\log{M_{\rm h,0}}\!\gtrsim\!12.5$ differ by no more than $\sim\!40\%$ at all times, meaning that both of these mechanisms contribute comparable power to sustaining turbulence. In particular, at redshift $z\!\sim\!2$, the difference between the energy injection rates from transport and accretion for galaxies in halos with $12.5\!\lesssim\log{M_{\rm h,0}}\!\lesssim14$ is less than $\sim\!20\%$. We conclude that, for a moderate level of accretion conversion efficiency, galaxies residing in haloes within the mass range of $12.5\lesssim\!\log{M_{\rm h,0}}\!\lesssim14$, corresponding to $\sim\!10^{12-13}\,\Msun$ at $z\!=\!2$, have their turbulence supported by transport and accretion with comparable power.

\smallskip We note that our model predicts that, at lower redshifts, the energy injection rates from accretion becomes larger in some of the galaxies. However, as discussed below, a large value of $\xi_a=0.6$ is not expected at these late times.

\item For a maximal accretion conversion efficiency, $\xi_a\!=\!1$ (bottom panel), discs in haloes with $\log{M_{\rm h,0}}\!\lesssim\!12$ still have their turbulence primarily driven by feedback throughout their lifetimes. In these discs, accretion is the secondary driver of turbulence, being more important than transport. More massive discs, residing in haloes with $\log{M_{\rm h,0}}\!\gtrsim\!12$, have their turbulence primarily driven by accretion throughout their lifetimes. In particular, at $z\!\sim\!4$, all discs have their turbulence primarily supported by accretion, with energy injection $\sim\!40\%$ larger than from transport. Feedback is the secondary driver in discs that reside in haloes with $12.5\!\lesssim\!\log{M_{\rm h,0}}\!\lesssim\!13$, while transport is the secondary driver for discs in haloes with $\log{M_{\rm h,0}}\!\gtrsim\!13$. As discussed below, such a high value of $\xi_a=1$ is expected at very high redshifts, but is not expected to be valid at low redshifts.
\end{itemize}

\smallskip For all of the cases studied above, we learn that when including all three drivers together, accretion can have a significant role in supporting the turbulence in massive discs, either as a primary driver, when $\xi_a=1$, as an equal power contributor, when $\xi_a=0.6$, or as a secondary driver, when $\xi_a=0.3$.
\subsubsection{Accretion, feedback \& transport - time varying $\xi_a$}\label{sec:xi_a_of_z}
The biggest uncertainty in our model is the conversion efficiency from accretion kinetic energy to turbulence in the disc, namely $\xi_a$. The value of this parameter represents the density contrast between the accreting stream and the disc, which is governed both by the ambient density and cross section of the stream, and by stream clumpiness. Since the value of this parameter is yet to be studied in a cosmological context, we treat it here in an ad-hoc way for a qualitative study.
\begin{figure*}
    \centering
    \includegraphics[width=1.5\columnwidth]{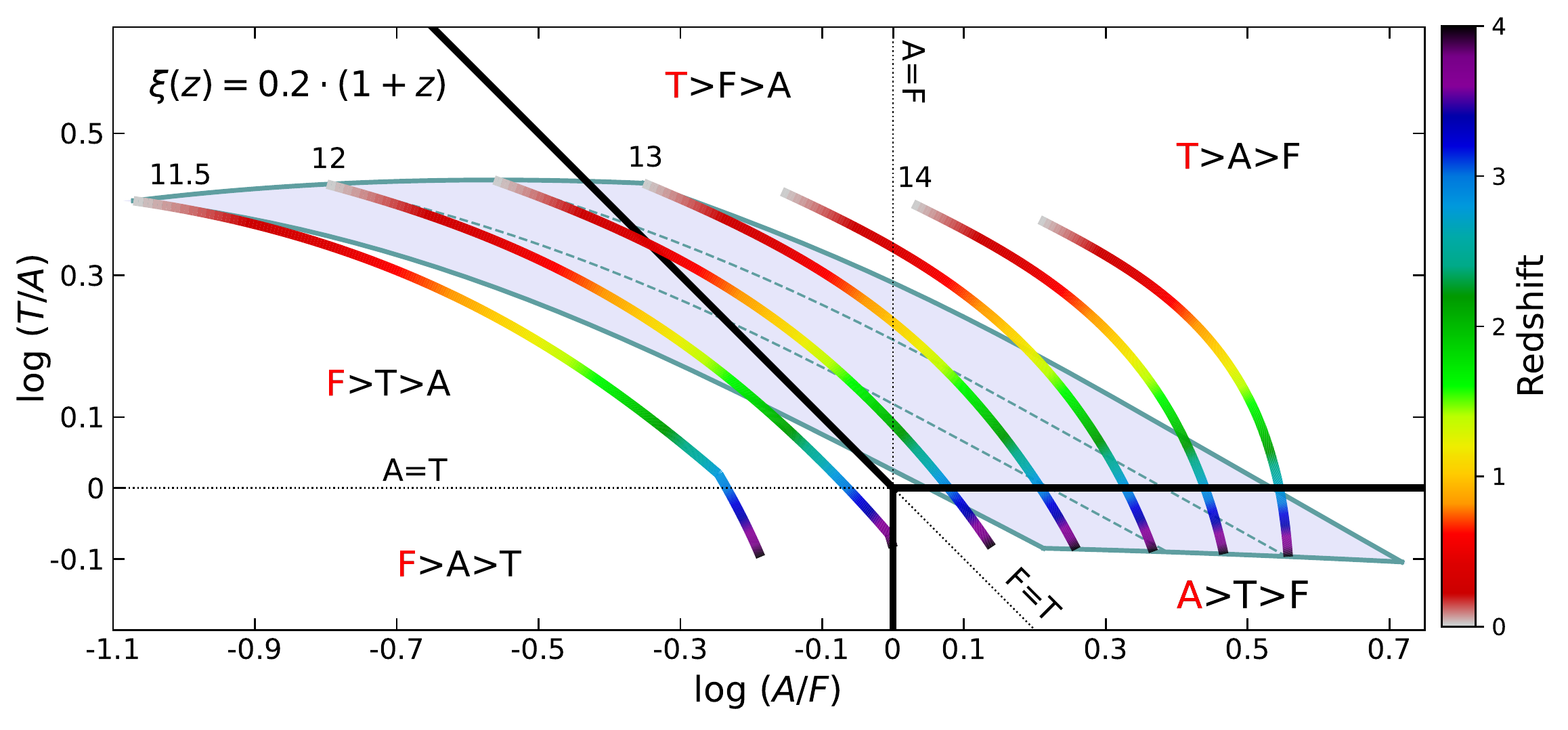}
    \caption{Same as figure \protect\ref{fig:accretion_tracks}, but with time varying $\xi_a$ (see text). We see that turbulence in massive discs transition from being primarily supported by accretion to transport. Low mass discs are primarily supported by feedback at low redshifts.}
    \label{fig:accretion_tracks_xi_varying}
\end{figure*}
\smallskip In particular, in the previous section we have treated it as fixed in time, but it is expected to decline in time following the stream clumpiness. For example, \cite{M18_GC} have studied the fragmentation of cosmological streams, and noted that the mass fraction in very-high density clumps that may serve as globular-cluster progenitors is a strong function of the halo mass and redshift, with streams being more clumpy at higher redshifts. We defer to future work a detailed analysis of the conversion efficiency of stream energy to disc turbulence as a function of halo mass and redshift. Here, for the purpose of a qualitative understanding, we limit ourselves to a potentially representative example of the time dependence of $\xi_a$, where we assume ad-hoc
\begin{equation}\label{eq:xi_a_of_z}
    \xi_a(z) = 0.2(1+z),\ z\leq4.
\end{equation}
This parameterization interpolates between a high value of $\xi_a$ at high redshifts, namely $\xi_a=1$ at $z=4$, to a low value at low redshifts, namely $\xi_a=0.2$ at $z=0$. The results of our model with this parameterization for $\xi_a$ are shown in figure \ref{fig:accretion_tracks_xi_varying}. While very crude, the figure shows qualitatively how turbulence support would evolve with an accretion efficiency that is declining with time. We read from the figure that at high redshifts, $z\gtrsim3$, turbulence in discs is mainly supported by accretion, at all halo masses at which the central galaxy is considered a disc (shaded region, see \S\ref{sec:Results}). The turbulence in discs becomes dominated by transport at $z\sim\!2.5$. Low mass discs, residing in halos with $\log{M_{\rm h,0}}\lesssim12$ today, have their turbulence primarily supported by feedback from the moment they become discs, and for the remainder their lifetimes. More massive discs remain in the regime where transport is the dominant driver of turbulence until $z=0$. From this crude approximation, we expect accretion to be a significant driver at high redshifts, where the streams are more likely to fragment into bound clumps. At low redshifts, the role of accretion weakens, and the evolution of disc turbulence is practically dominated by feedback and transport (\S\ref{sec:feedback_and_transport}).

{It is worth evaluating the qualitative effects on the transport rate of values of $Q$ below unity, as expected for thick discs in marginal instability, $Q\!\sim\!0.67$ \citep{Goldreich65}. For our fiducial $n=1$, $\sigma_{\rm SN}$ is independnt of $Q$ (eqs. \ref{eq:ff_to_td},\ref{eq:sigma_sn}), while $\sigma_{\rm acc}\propto Q^{2/3}$ (eq. \ref{eq:sigma_acc}) and $F\propto Q^{-1}$ (eq. \ref{eq:F_of_sigma}). Combined, a lower value of $Q$ yields a higher transport rate, which results in a lower disc mass, and therefore a smaller $\sigma_g$. A smaller velocity dispersion increases the contribution by accretion and feedback at the expense of transport. For the constant $\xi_a=0.3$ this makes no qualitative difference, however for $\xi_a=0.6$, accretion becomes dominant by $z\!\sim\!3$ (and below), compared to our fiducial results. For the time-varying $\xi_a$ model, the transport becomes dominant over the accretion only at $z\!\sim\!1$ and below, compared to our fiducial results.}

\subsection{Shutoff of transport}\label{sec:shutoff_of_transport}
As mentioned in \S\ref{sec:QE_turb} and in Appendix \ref{sec:sigma_c_app}, there is a critical velocity dispersion $\sigma_c(z,M_{\rm h})$ below which instability driven inflow shuts off in order to maintain the energy equilibrium (eqs. \ref{eq:QE_enc}, \ref{eq:QE_visc}). While we have seen that in all models considered in \S\ref{sec:Results} the transport always has some contribution, it is interesting to see how the addition of accretion driven turbulence changes the evolution of turbulence in galaxies when transport is absent.

\smallskip In such cases, $\alpha_c=0$ or $\tau=0$ for the two different methods of computing the transport rate, respectively, namely the disc no longer hosts clumps and turbulent viscosity is no longer exerting torques. Equations \ref{eq:QE_enc} and \ref{eq:QE_visc} reduce to the same equation
\begin{equation}\label{eq:QE_no_trans}
\begin{split}
    \frac{2}{3}\epsff\pms\chi^{-1}\sigma_{\rm g}^{-1} + &\frac{1}{3}\xi_a\frac{G\dot{M}_{\rm g,acc}Q}{f_{\rm g,Q}\sqrt{2(1+\beta)}}\sigma_{\rm g}^{-3}\\
    &   = \sqrt{2(1+\beta)}\gamma_{\rm diss}^{-1}\lbrac{\frac{Q}{f_{\rm g,Q}}}^{-n}.
\end{split}
\end{equation}
Since $\xi_a$ and $\dot{M}_{\rm g,acc}$ refer to external drivers, they are not self regulated by the disc. The momentum per unit of stellar mass formed, $\lbrac{p/m}_*$, is determined by the internal physics of supernova explosions, and hence it is also not self regulated by the disc. This leaves $\chi,\sigma_{\rm g},\epsff$ and $Q$ as possible disc properties that can be self regulated. The parameter $\chi$ depends on $Q$ and on other parameters that are regulated by the assumed vertical force balance, and are unrelated to the unstable state of the disc (see eq. \ref{eq:ff_to_td}). The variations in these parameters are not expected to introduce values beyond the range of values already considered in this work. Hence, the only independent variables that can be self-regulated, in the case of no transport, are $Q,\epsff,\sigma_{\rm g}$.

\smallskip When no transport is in effect, the solution to eq. \ref{eq:mass_conservation} approaches the approximate solution
\begin{equation}\label{eq:bathtub_ss}
    \dot{M}_{\rm g,acc} = \dot{M}_{\rm SF} = \epsff\frac{\Mg}{t_{\rm ff}}.
\end{equation}
Plugging this expression in eq. \ref{eq:QE_no_trans}, we get
\begin{equation}
\begin{split}
    \frac{2}{3}\epsff\pms\chi^{-1}\sigma_{\rm g}^{-1} &+ \frac{1}{3}\xi_a\epsff\frac{G\Mg Q}{f_{\rm g,Q}\chi\td\sqrt{2(1+\beta)}}\sigma_{\rm g}^{-3}\\ & =\sqrt{2(1+\beta)}\gamma_{\rm diss}^{-1}\lbrac{\frac{Q}{f_{\rm g,Q}}}^{-n}.
    \end{split}
\end{equation}
From the definition of the Toomre-$Q$ parameter (eq. \ref{eq:Q_parameter}), we have that $G\Mg\Qg = \sqrt{2(1+\beta)}\Vd\Rd\sigma_{\rm g}$. Plugging this into the above equation, we get
\begin{equation}\label{eq:QE_steady_state_no_trans}
    \frac{2}{3}\epsff\pms\chi^{-1}\sigma_{\rm g}^{-1} + \frac{1}{3}\xi_a\epsff\chi^{-1}\lbrac{\frac{\sigma_{\rm g}}{\Vd}}^{-2} = \sqrt{2(1+\beta)}\gamma_{\rm diss}^{-1}\lbrac{\frac{Q}{f_{\rm g,Q}}}^{-n}.
\end{equation}
Since $\chi\propto Q$ (eq. \ref{eq:ff_to_td}), for our fiducial value of $n=1$, this equation is independent of the value of $Q$, meaning that even if $Q$ is free to vary, it will not be set by the energy conservation.

\smallskip We first let $\epsff$ be fixed. Under the steady state approximation, eq. \ref{eq:bathtub_ss}, the solution of eq. \ref{eq:QE_steady_state_no_trans} is $\sigma_c$ of eq. \ref{eq:sigma_c} for any value of $Q$, i.e. in the case of a fixed $\epsff$, $\sigma_g$ remains at the level of $\sigma_c$. When only feedback is considered and the contribution of accretion to generating turbulence is ignored, we have $\sigma_{\rm c} = \sigma_{\rm SN}$, independent of mass. However, when accretion is included, the critical value of turbulence sustained by feedback and accretion in the absence of transport depends on both halo mass and redshift, $\sigma_{c}(M_{\rm h},z)$ from \ref{eq:sigma_c}.

\smallskip On the other hand, if $\epsff$ can be self regulated by the disc while $Q$ remains fixed, we can solve for $\epsff$ from eq. \ref{eq:QE_steady_state_no_trans}, to get
\begin{equation}\label{eq:QE_epsff}
    \epsff = \chi\sqrt{2(1+\beta)}\gamma_{\rm diss}^{-1}\lbrac{\frac{Q}{f_{\rm g,Q}}}^{-n}\frac{\sigma_{\rm g}}{\frac{2}{3}\pms+\frac{1}{3}\xi_a\Vd^2/\sigma_{\rm g}}.
\end{equation}
For a given $\sigma_{\rm g}$, this is smaller than the prediction of eq. 54 in \cite{K18}, as we would expect: since we have added an extra source of energy, in the form of accretion, we require less energy input from star formation, and thus require less extreme values of $\epsilon_{\rm ff}$. Note that for our fiducial $n=1$, $\epsff$ is independent of the value of $Q$, since $\chi\propto Q$. Equation \ref{eq:QE_epsff} suggests that, when transport is absent and the star formation efficiency is allowed to vary, more turbulent discs have a higher star formation efficiency. 

\begin{figure*}
    \centering
    \includegraphics[width=2\columnwidth]{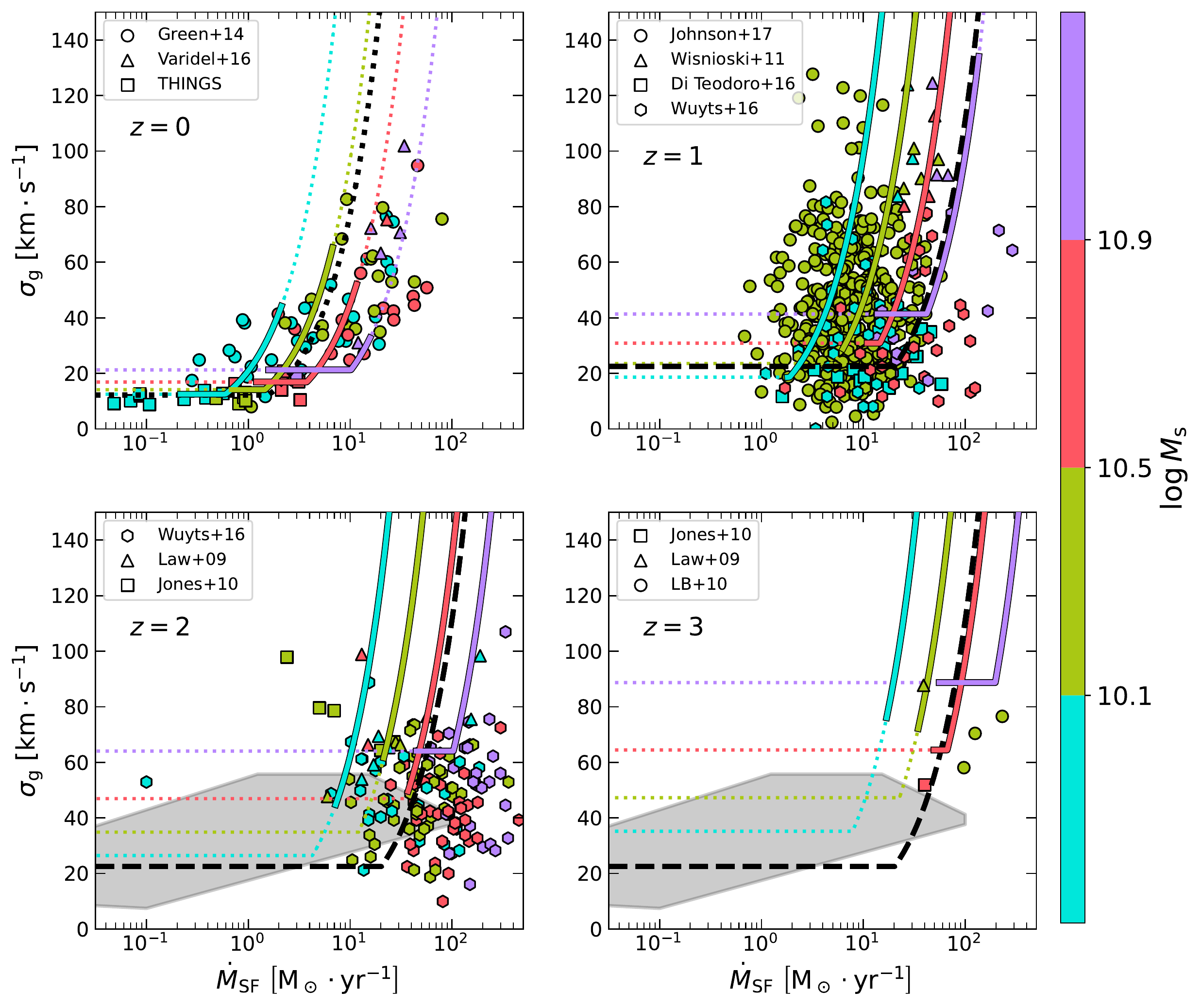}
    \caption{The relation between the gas velocity dispersion, $\sigma_{\rm g}$, and the star formation rate, $\dot{M}_{\rm SF}$. Each panel refers to a different redshift. Colors represent different stellar mass bins at each redshift. The colored curves in each panel represent the relation as predicted by our model, for fixed $M_{\rm h}(z) = 10^{11.5}, 10^{12},10^{12.5},10^{13}\,\Msun$ at each redshift, converted into stellar mass using the stellar-to-halo mass relation by \protect\cite{Behroozi13}. The solid section of each curve indicates the range of SFRs within ${\rm 0.5\, dex}$ of the value at the main sequence, taken from \protect\cite{Whitaker2012}. The dashed and dotted lines show the results of the high-z and local spiral model of \protect\cite{K18}, respectively. The shaded regions are the results from isolated simulations (with feedback) by \protect\cite{Ejd}, high gas fraction runs. Symbols represent observational results of ${\rm H\alpha}$ and ${\rm HI}$ at low redshifts \protect\citep{Leroy08,Walter08,Ianj12,Green14,Varidel16} and high redshifts \protect\citep{Law09,Jones10,LB10,Wisnioski11,DiTeodoro16,Wuyts16,Johnson17}. The masses assigned by color to the symbols of \protect\cite{Jones10}, \protect\cite{DiTeodoro16} and \protect\cite{Johnson17} are the median masses of each sample. We see a general agreement between our model and the observations, both in terms of the level of turbulence and the mass trends.}
    \label{fig:sig_vs_Sfr}
\end{figure*}
\subsection{$\dot{M}_{\rm SF}-\sigma_{\rm g}$ relation}
We can use our model to calculate the $\dot{M}_{\rm SF}-\sigma_{\rm g}$ relation from eqs. \ref{eq:Q_parameter} and \ref{eq:sf_law}. According to our model, $\dot{M}_{\rm SF}\propto \sigma_{\rm g}$, whenever $\sigma_{\rm g}\geq \sigma_{\rm c}\left(M_{\rm h},z\right)$. In figure \ref{fig:sig_vs_Sfr}, we show the $\dot{M}_{\rm SF}-\sigma_{\rm g}$ in our model\footnote{We note that the curves are not produced as a result of integration over the history of a galaxy, but rather represent the functional relation between $\dot{M}_{\rm SF}$ and $\sigma_{\rm g}$.} (assuming the time-varying parameterization for $\xi_a(t)$\footnote{{While the current tentative choice of the time evolution of $\xi_a$ is ad-hoc, the dependence of $\sigma_c$ on $\xi_a$ is rather weak, a power law with a power of $1/3$ (eqs. \protect\ref{eq:sigma_acc} and \protect\ref{eq:sigma_c}), as seen in figure \protect\ref{fig:app_sigma_c}. The redshift evolution of the $\dot{M}_{\rm SF}-\sigma_g$ relation is therefore not very sensitive to the choice of $\xi_a(z)$.}}, eq. \ref{eq:xi_a_of_z}) at four different redshifts and for four different masses. Each curve in each redshift is for a fixed halo mass at that particular redshift, converted into stellar mass using the stellar-to-halo mass relation, as parameterized in eq. 3 of \cite{Behroozi13}. The solid section of each curve indicates the range of SFRs within ${\rm 0.5\, dex}$ of the value at the star forming main sequence, fitted by \cite{Whitaker2012} to galaxies at $0\!\lesssim\!z\!\lesssim\!2.5$.

When $\sigma_{\rm g}=\sigma_{\rm c}$, transport shuts off, and the turbulence remains at the level of $\sigma_{\rm c}$ (see discussion in \S\ref{sec:shutoff_of_transport}). Our model predicts variations in $\sigma_{\rm g}$ as a function of mass at a fixed $\dot{M}_{\rm SF}$, due to the dependence of $\sigma_c$ on the halo mass. This is because galaxies that reside in more massive haloes undergo stronger accretion (due to our chosen accretion model, eq. \ref{eq:gas_accretion_rate}) and therefore have a larger $\sigma_c$ (see Appendix \ref{sec:sigma_c_app})\footnote{We can think of other sources of variations in $\sigma_c$, which we do not address in this paper. One source is variations about the average cosmological accretion rate which are translated to variations in $\sigma_c$, and thus cause variations in $\sigma_{\rm g}$ at a fixed mass. Another source is variations in the strength of feedback at a fixed SFR. Such variations could arise, for example, from variations in $(p/m)_*$ or $\epsff$ or from variations in the clustering of supernova \citep{Gentry17}.}.
The results of analytical models by \cite{K18} are also shown in figure \ref{fig:sig_vs_Sfr} (their `local spirals' and `high-$z$' models). The general behavior is similar between our model and theirs, with the floor in $\sigma_g$ in the models by \cite{K18} generally smaller, due to the absence of accretion in their model.

In figure \ref{fig:sig_vs_Sfr} we compare the relation predicted by our model to ${\rm H\alpha}$ observations in high redshifts and ${\rm H\alpha}$ and ${\rm HI}$ observations at low redshifts (see the figure caption). We use a subset of the observational results compiled in Appendix B of \cite{K18}. We can see a general agreement between our model and the observations\footnote{The galaxies observed by \cite{Green14} and \cite{Varidel16} usually lie above the main sequence, being selected intentionally as local analogs of high redshift star forming galaxies. Even though adopting our model well above the main sequence at $z=0$ is in agreement with these samples, we note that adopting the average cosmological values for our model parameters for comparing with these galaxies may be incorrect.}, with the mass trend in the same ballpark (mass bins indicated by the color), as well as the varying levels of $\sigma_{\rm g}$ at a given SFR. We note, however, that the high levels of $\sigma_g$ at high SFRs are usually observed for ${\rm H\alpha}$ rather than atomic or molecular gas, a distinction we have not made in our model.

\cite{Ejd} have studied this relation in simulations of isolated galaxies, and found that when looking at the gas as a whole, the high levels of $\sigma_g$ observed at high redshifts are not reproduced. On the other hand, the ${\rm H\alpha}$-emitting gas alone does seem reproduce the high values for $\sigma_{\rm g}$. \cite{Ejd} have further argued that the high levels of turbulence could be an observational artifact, as a result of beam smearing and inclination effects. While their fiducial sample was not able to reproduce these high levels of turbulence (see shaded regions in figure \ref{fig:sig_vs_Sfr}), when considering these aforementioned effects, they were able to achieve high values. We note, however, that these simulations, not being in a cosmological setting, do not take into account the possible driving of turbulence by accretion, which can potentially strengthen the turbulence for the total gas.
\section{Discussion}\label{sec:discussion}
It is worth discussing the simplifying assumptions of our model, potential caveats and possible improvements in future work.
\subsection{Spatial variations}
While we have approximated the galaxy properties (e.g. star formation, mass transport) as spatially-independent and only evolving with time, this is clearly an over-simplification. While some of the processes discussed here do approach a spatially-independent solution in a steady state \citep{KB10,Forbes12}, star formation, feedback and accretion may depend on the position inside the galaxy.

\smallskip It has been shown that the cold streams that feed the disc are usually confined to a plane, which is not necessarily parallel to the disc plane \citep{Danovich12}. While the streams tend to co-rotate with the disc, about a third of the mass is in-streaming counter rotating to the disc \citep{Danovich15}. However, the detailed profile of how streams feed the galaxy at different radii has not been studied thoroughly. At $z=0$, \cite{Trapp21} used the FIRE simulations \citep{Hopkins14} to show that in-streaming gas tends to pile up at the outskirts of the galaxy, and does not directly contribute much flux in inner radii. Since high redshift galaxies tend to reside in more messy regions, and to be fed by more intense inflowing cold streams \citep{DB06,Dekel09}, it is not obvious at all that this picture remains valid at high redshifts.

\smallskip In future work, we plan to study in detail the different profiles of the different sources and sinks, starting from the feeding of mass and angular momentum by streams at different radii of the disc. These will be used as boundary conditions for a more detailed, spatially dependent model.
\subsection{Outflows}\label{sec:outflows}
Besides stirring up turbulence, feedback is also responsible for generating outflows from the galaxy. This can be due to two main mechanisms. One is supernova explosions and stellar winds, which are related to the SFR of the galaxy, and are effective in low mass galaxies, below the critical mass of $\log{M_{\rm h}}\sim 12$ \citep{DekelSilk86}. The second is due to AGN feedback, which becomes effective after the galaxies go through a wet compaction, above the same critrical mass \citep{Lapiner21}. 

\smallskip The strength of outflows is usually characterized by the mass loading factor, $\eta = \dot{M}_{\rm g,out}/\dot{M}_{\rm SF}$, where $\dot{M}_{\rm g,out}$ is the mass loss rate due to outflows. Observationally, $\eta$ varies in the range $\eta\sim0.1\!-\!1$ for star forming galaxies in the redshift range $0.6\!<\!z\!<\!2.7$ \citep{Davies19,Forster19}, while simulations show a wider range of values, $\eta\sim 1-100$ \citep{Muratov15,Nelson19}, with or without AGN \citep{Mitchell20}, for galaxies in the same redshift range. The discrepancy between the values of the mass loading factors in different simulations lies in the differences between the assumed feedback models. These feedback models are usually fine tuned to match the observed stellar-to-halo mass relation. However, this practice often disregards other properties of the galaxies that are also affected by feedback. For example, the VELA simulations \citep{Ceverino14} produce long lived giant clumps with abundances and physical properties consistent with observations \citep{M17, G18, Ginzburg21, Dekel21}. However, these simulations often produce galaxies that are more massive than observed, due to the relatively weak feedback model they assume. On the other hand, simulations that implement models with stronger feedback, in order to better agree with the observed stellar-to-halo mass relation, often fail to produce the abundance of long lived clumps \citep{Genel12b,Oklopcic2017}. This conflict between the role of feedback on galactic and clump scales needs to be addressed in future work, which calibrates a feedback model that obeys both large- and small-scale constraints.

\smallskip Outflows should introduce another sink term in eq. \ref{eq:mass_conservation}, of the form $\eta\dot{M}_{\rm SF}$. While it has been shown that the mass outflows from stellar feedback can have an important effect on the evolution of the galaxy \citep{DM14}, this has been studied disregarding other potential forms the energy released by stellar feedback can take, e.g. the turbulent form. Hence, the mass loading factor needs to be modeled consistently, taking into account that not all of the energy from the stellar feedback goes into outflows. \cite{HH16} have developed a turbulent model for the mass loading factor, however their model takes into account only turbulence driven by supernova feedback while disregarding the other potential drivers of turbulence considered here. Their model needs to be modified to allow for multiple drivers of turbulence, in order to consistently estimate the mass loading factor, which is beyond the scope of this paper.

{In order to crudely evaluate the possible qualitative effect of outflows on our results, at the risk of not being self-consistent, we added to eq. \ref{eq:mass_conservation} a sink term of the form $\eta\dot{M}_{\rm SF}$. The direct effect of a constant mass loading factor is an overall reduction of the mass of the disc. This in turn reduces the velocity dispersion of the disc, thus bringing $\sigma_{\rm g}$ closer to $\sigma_c$, which means that the role of transport in balancing turbulence dissipation becomes less important. For the fixed $\xi_a$ models, a value of $\eta=1$ keeps the transport dominant in high mass discs for $\xi_a=0.3$, but for $\xi_a=0.6$, the outflows make the accretion dominant already at $z\!\sim\!3$ and below, compared to $z\!\sim\!1$ with no outflows. For the time-varying $\xi_a$ model, the transport overtakes the accretion only at $z\!\sim\!1$ and below, compared to $z\!\sim\!3$ and below with no outflows. A self-consistent analysis incorporating outflows is deferred to future work.}
\subsection{Recycling}
While the external gas accretion is the main source of gas to the galaxy, galaxies that undergo outflows are also subject to recycling of this previously outflowing gas that was unable to escape from the halo. \cite{DM14} have proposed that recycling may be important to explain discrepancies between theoretical models of galaxy formation and observations at around $z\!\approx\!2$, after substantial star formation has occurred. Also, numerical simulations have shown that recycled winds can significantly contribute to the overall accretion of gas, and can sometimes dominate the supply of gas to the galaxy \citep{Alcazar17}. However, there is still disagreement between numerical simulations about the relevant timescale for recycling, and it can vary between $0.2\!-\!1\,\Gyr$ \citep{Oppenheimer10,Alcazar17,Tollet19}, and some simulations find the recycling timescale to be longer than the Hubble time \citep{Mitchell20b}.

\smallskip Modelling recycling theoretically is a challenging task. \cite{DM14} modeled recycling simply as a lower value of the mass loading factor for outflows, however this treatment assumes that the recycling is instantaneous, namely that the timescale for recycling is much shorter than other dynamical timescales in the problem. How recycling drives turbulence is also unknown, in particular its effect on $\xi_a$ and $\dot{M}_{\rm g,acc}$, and requires more numerical work to inspire analytical models. Recycling is therefore an open question, that we do not attempt to address here.
\subsection{Modeling of accretion driven turbulence}
Our implementation of accretion-driven turbulence due to the incoming streams has been extremely crude. First, the parameterization of the energy deposited in turbulence by accretion is an over-simplification. The accretion occurs along narrow streams, and the location of the collision between the streams and the disc is not known. Second, the conversion efficiency of kinetic energy associated with gas collisions to turbulence was either assumed constant throughout the evolution of the galaxy, or was parameterized ad-hoc to have a qualitative idea of how its evolution with time affects the disc as a whole. \cite{KH10} suggest that the conversion efficiency is $\xi_a\!\sim\!\Delta$, where $\Delta$ is the density contrast between colliding flows. \cite{M18_GC} have estimated analytically that the clumpiness of the streams is a function of both the stream's host halo mass and redshift, so one would generally expect $\xi_a$ to be a function of those two parameters. In future work, we plan to use the models proposed by \cite{M18_GC,M20} to estimate analytically the evolution of the conversion parameter. Finally, due to the rotation of the disc, the orientation at which the streams hit the disc can affect the conversion of kinetic to turbulent energy, e.g. streams co-rotating with the disc will stir less turbulence than counter rotating streams. In future work, we plan to study analytically and using simulations how accretion in general, and clumpy accretion in particular, drives turbulence in the disc.
\subsection{Mass transport}
In this paper, we computed the rate of instability-driven mass transport within the disc in two alternative ways - encounters between clumps and turbulent viscous torques. We found that, when imposing an energy equilibrium, these two processes produce the same mass transport rate, with a non trivial dependence on the velocity dispersion. However, these two processes do not necessarily cover the whole range of processes involved. \cite{DekelRings} have found that once the galaxy exhibits a massive bulge, it tends generate a long-lived extended ring with suppressed mass transport rate. The transport timescale in this case was found to be much longer than the timescale for mass transport in a VDI disc. The timescale for clump migration as estimated by dynamical friction within the disc has been found to be comparable to and slightly shorter than the estimate based on encounters \citep[][Appendix A]{Dekel13}. It is still not clear what is the dominant mass transport mechanism, and there is more theoretical work to be done, both analytical and using simulations, to better understand the nature of mass transport in the disc and the implied rate. We plan to study these in more detail in future work.

Furthermore, mass that is being transported down the potential well eventually ends up in the bulge. This induces a stabilizing effect, as described in \cite{DekelRings}. Furthermore, the gas that flows into the bulge feeds the supermassive black hole \citep{Lapiner21}, induces AGN feedback, hence another potential source of turbulence we have not considered here.

\section{Conclusions}\label{sec:conclusions}
In this paper, we have addressed the evolution of turbulent support in disc galaxies, as a function of mass and redshift, assuming that they are self-regulated in a marginal Toomre disc instability, with the appropriate level of turbulence maintained for many dynamical times simultaneously by three energy sources that balance the dissipative losses. These energy sources are stellar feedback, mass transport within the disc, and cosmological accretion onto the disc. This is a generalization of previous models which dealt with only two of the driving sources at a time. We used an analytical bathtub model for the evolution of the discs, based on the following assumptions:
\begin{itemize}
    \item Galaxies are modeled as discs of gas and stars in a turbulent, marginally Toomre unstable state, embedded in dark matter haloes.
    \item The discs smoothly transition from gas-dominated discs at high redshifts to two component discs of gas and stars at low redshifts
    \item Discs gain fresh gas from external accretion at the average cosmological rate.
    \item Discs lose gas by either star formation or mass transport through the disc to the central bulge.
    \item The transport rate has been computed in two alternative ways, via clump encounters and via turbulent viscous torques, with very similar results.
    \item Turbulence in the discs is driven in concert by three energy sources, namely, supernova feedback, instability-driven torques and the associated mass transport and the impact of clumpy accretion onto the disc.
    \item The intensity of accretion-driven turbulence is characterized here by an energy conversion efficiency parameter, and we considered low, moderate and maximal efficiencies.
    \item Turbulence dissipation is balanced by the three energy sources, resulting in an energy balance in the disc.
\end{itemize}

\smallskip We then numerically solve a conservation equation for the gas mass and turbulence energy, each time considering a different set of turbulence drivers and eventually considering them all in concert. Under the assumptions of turbulent energy balance and marginal Toomre instability, this directly allows us to compute the turbulent state of the disc, and to determine which turbulence driver is the dominant one, for the given halo mass and redshift. Our results are as follows:
\begin{itemize}
    \item When including only feedback and transport as drivers, turbulence is mainly supported by transport at $z\!>\!1.5$, for all disc masses. At lower redshifts, discs that reside in haloes with masses that evolve to ${M_{\rm h}(z=0) = M_{\rm h,0}\!\lesssim\!10^{12}\,\Msun}$ today, become mainly supported by supernova feedback.
    \item When including only accretion and transport as drivers, we find that different levels of stream energy conversion efficiencies produce qualitatively different results. For a low conversion efficiency, turbulence in all discs, at all times and masses, is primarily supported by transport, at the level of $\sim\!70\%$. For a moderate conversion efficiency, transport and accretion contribute comparable power to sustaining turbulence. Lastly, for a maximal conversion efficiency, $70\%$ of the turbulence dissipation is supported by accretion.
    \item When considering all three drivers together, we find that for all levels of conversion efficiency, discs that reside in haloes that evolve into $M_{\rm h,0}\!\lesssim\!10^{12}\,\Msun$ at $z=0$ have their turbulence primarily supported by supernova feedback throughout their lifetimes. For discs in haloes more massive than $10^{12.5}\,\Msun$ today, the result depends on the accretion conversion efficiency as follows:
    \begin{enumerate}
        \item  For a low conversion efficiency, below $\sim\!40\%$, the turbulence is primarily supported by transport, throughout the galaxy lifetime.
        \item For a moderate accretion efficiency, at $40\%-70\%$, the turbulence is driven primarily by transport at high redshifts, however the contribution of transport is at most $40\%$ larger than that of accretion. In particular, at $z\!\sim\!2$, the difference between the contributions of transport and accretion to sustaining turbulence is at most $20\%$. We thus conclude that accretion and transport contribute to sustaining turbulence with comparable power for these galaxies.
        \item For a maximal accretion conversion efficiency, close to unity, the primary driver of turbulence is accretion at all times in the galaxy's history. In particular, at $z\!\sim\!4$, accretion dominates by more than $40\%$ over transport.
    \item Finally, in order to study how an $\xi_a$ that declines with time qualitatively affects the evolution, we make a very crude approximation for its evolution as a function of redshift. We find that discs within halos that end up with with $M_{\rm h,0}\lesssim10^{12}\,\Msun$ have their turbulence supported by feedback at all times, while turbulence in discs within halos of $M_{\rm h,0}\gtrsim10^{12.5}\,\Msun$ transition between a marginal accretion dominance to transport dominance at around $z\!\sim\!3$, when their host halo's masses were $M_{\rm h}(z=3)\!\gtrsim\!10^{11.5}\,\Msun$.
    
    \item The main lesson is that, even though different drivers dominate the turbulence at different stages of the disc's evolution, the relative roles of the three energy sources are comparable to within a factor of $2-3$ in all disc galaxies at all times.
    \end{enumerate}
     \item Our model predictions for the $\dot{M}_{\rm SF}-\sigma_{\rm g}$ relation are largely consistent with the observations both at low and high redshifts. They are also crudely consistent with simulations of isolated galaxies for low values of SFR, though they may predict for the whole gas higher values of $\sigma_g$ for high SFR at high redshifts.
\end{itemize}

Our simplified model suggests that accretion driven turbulence plays an important and sometimes the dominant role in sustaining turbulence in galactic discs, in parallel with transport and supernova feedback. In the current version of the model, the three ingredients were based on simplified parametric expressions. This sho uld be improved next, based on more detailed physical modeling of the driving of turbulence by accretion, the instability-driven transport rate, and the way feedback drives turbulence. Another important generalization would be to consider radial variations of the physical quantities within the disc, and demanding local equilibrium.

\section*{Acknowledgements}
This work has been supported by the grants ISF 861/20 and DIP 030-9111 and by a Milner Fellowship to OG.
NM acknowledges support by the Israel Science Foundation (grant No. 3061/21).
MRK acknowledges support from the Australian Research Council through its Future Fellowships funding scheme, award FT180100375.

\section*{Data availability}
Data and results underlying this article will be shared on reasonable request to the corresponding author.



\bibliographystyle{mnras}
\bibliography{example} 




\appendix
\section{Calculating $\sigma_c$}\label{sec:sigma_c_app}
\begin{figure}
    \centering
    \includegraphics[width=\columnwidth]{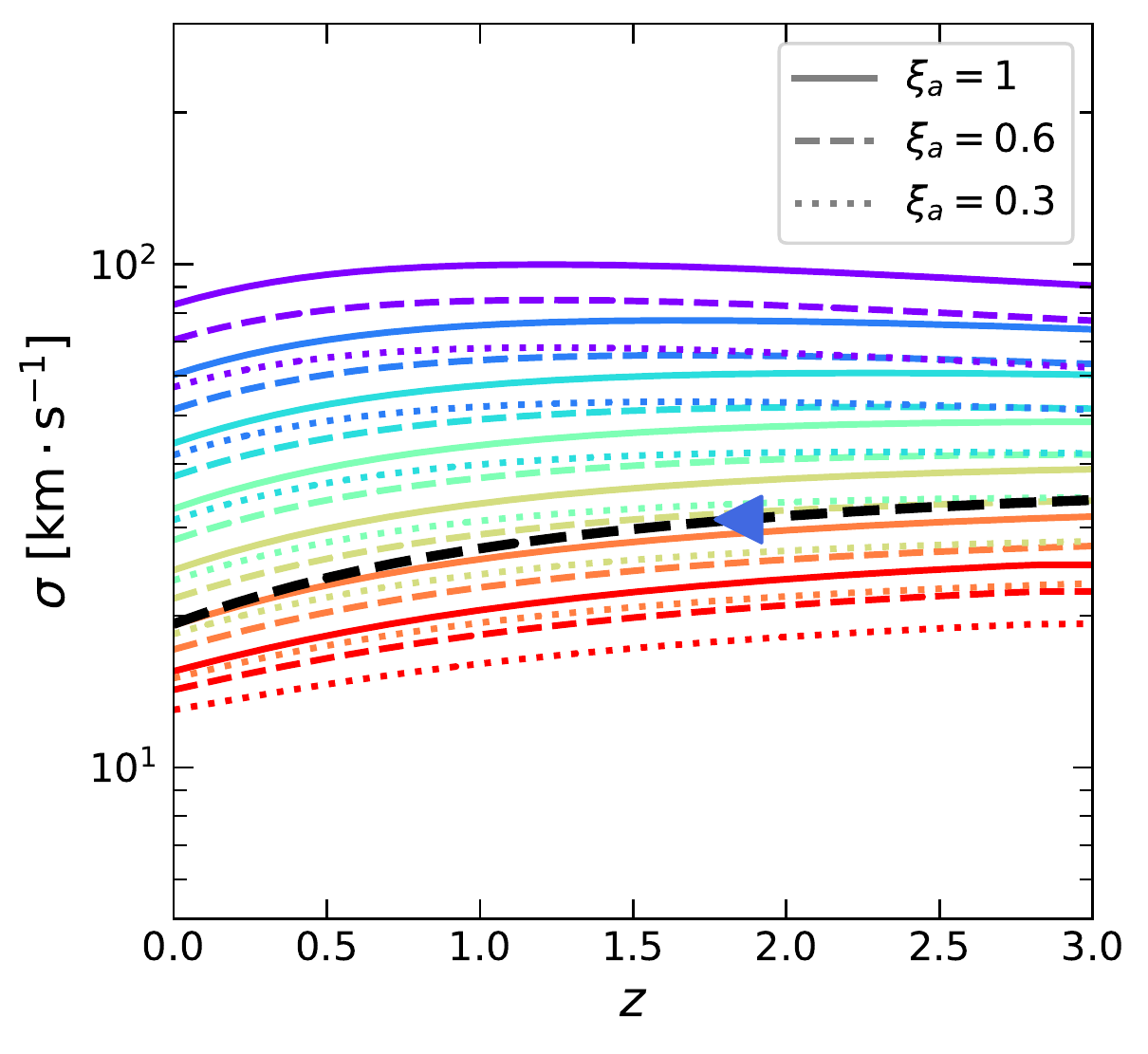}
    \caption{The evolution of the critical velocity dispersion, $\sigma_c$, below which mass transport shuts off. Colored lines show the evolution of $\sigma_c$ for haloes with $\log{M_{\rm h,0}}=11.5-14.5$, in intervals of $0.5\,{\rm dex}$. Solid, dashed and dotted lines show the evolution for $\xi=1,0.6,0.3$, respectively. The black dashed line shows the evolution of the velocity dispersion in a disc with $\log{M_{\rm h,0}}=12$, evaluated using $\xi_a=0.6$. The relevant $\sigma_c$ curve for this galaxy is the dashed orange curve. The blue arrow indicates the point in time in which this galaxy becomes a disc. We can see that the galaxy is always above this critical value, indicating the mass transport is always present, though its contribution decreases with time.}
    \label{fig:app_sigma_c}
\end{figure}

The critical velocity below which gravitationally driven mass transport halts is calculated by solving a cubic equation of the form
\begin{equation}\label{eq:cubic_eq}
    1-\frac{A}{x}-\frac{B^3}{x^3}=0.
\end{equation}
In our case, $x=\sigma_c$, $A=\sigma_{\rm SN}$ and $B=\sigma_{\rm acc}$ (see eqs. \ref{eq:SR_both}). Since $\sigma_{\rm acc}$ depends on $\dot{M}_{\rm g,acc}$, which in turn depends on $M_{\rm h}$, $\sigma_c$ will be a function of halo mass as well as redshift. The cubic in eq. \ref{eq:cubic_eq} is monotonically increasing, and therefore has a unique real root. It is given by
\begin{equation}\label{eq:sigma_c}
    \sigma_c = \frac{1}{3}\lbrac{\sigma_{\rm SN}+\frac{\sigma_{\rm SN}^2}{C^{1/3}}+C^{1/3}},
\end{equation}
where
\begin{equation}
    C = \frac{1}{2}\lbrac{2\sigma_{\rm SN}^3+27\sigma_{\rm acc}^3+3\sqrt{12\sigma_{\rm SN}^3\sigma_{\rm acc}^3+81\sigma_{\rm acc}^6}}.
\end{equation}
Figure \ref{fig:app_sigma_c} shows the evolution of $\sigma_c$ for different $\log{M_{\rm h,0}}$, for the three levels of conversion efficiencies considered. As the conversion efficiency increases, the values of $\sigma_c$ increase, indicating that higher levels of turbulence can be sustained by feedback and accretion, without the need for mass transport (see discussion in \S\ref{sec:shutoff_of_transport}). 

As discussed in the text, discs in our model never reach this threshold of $\sigma_c$. As an illustration, the black line in figure \ref{fig:app_sigma_c} shows the gas velocity dispersion of a disc with $\log{M_{\rm h,0}}=12$, evaluated using a model with $\xi_a=0.6$. The value for $\sigma_c$ which correspond to this halo is the orange dashed line. We see that the velocity dispersion of this disc (dashed black curve) is always larger than $\sigma_c$ (dashed orange curve), meaning transport always occurs in this disc.

\section{Convergence to a steady state solution}\label{sec:QSS_app}
\begin{figure}
    \centering
    \includegraphics[width=\columnwidth]{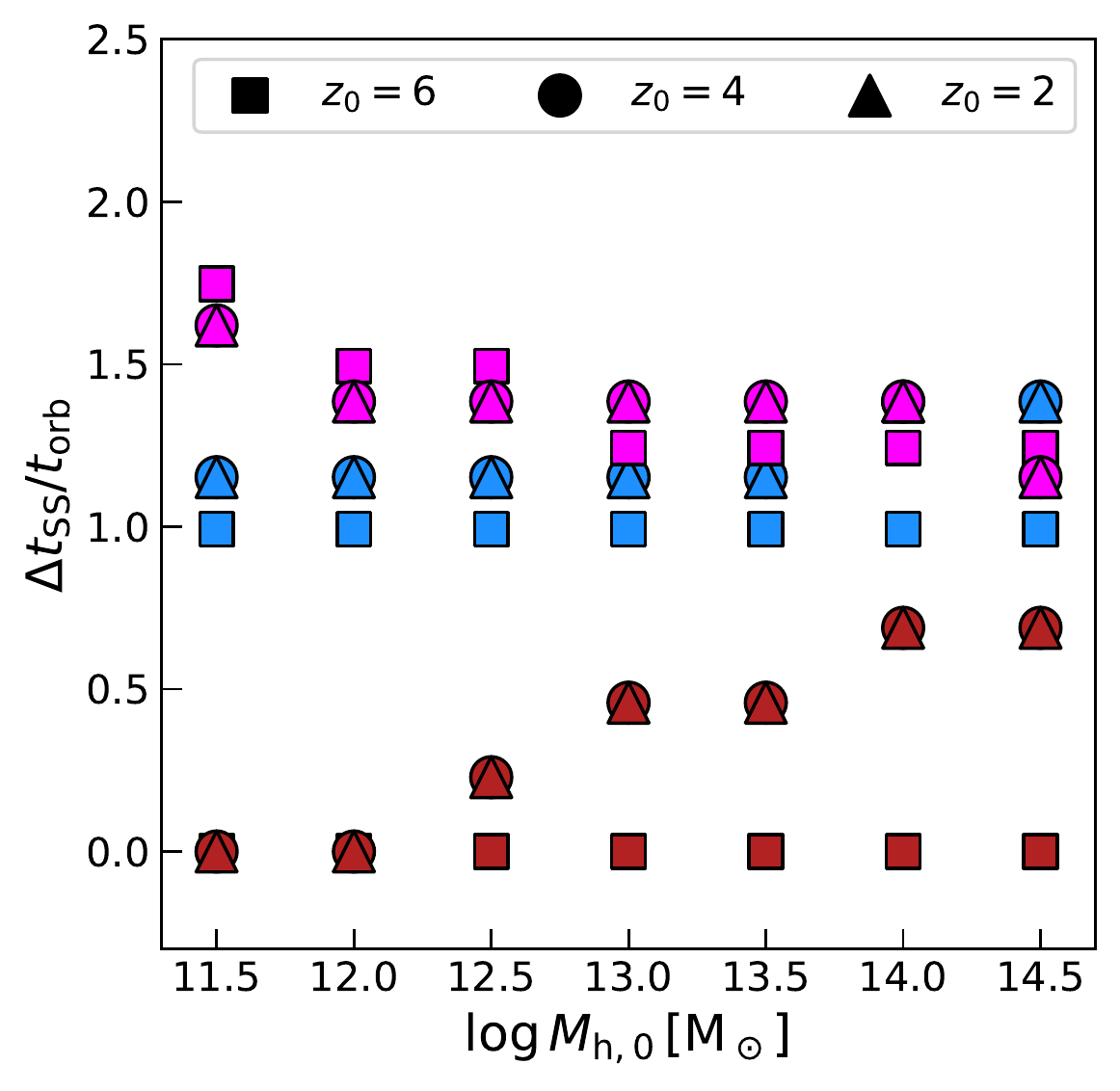}
    \caption{The convergence time of the numerical solution to the steady state unique solution. Shown are the number of disc orbital times until the solution converges to the steady state solution, for different values of $\log{M_{\rm h,0}}$, for different initial conditions and initial redshifts. The different symbols indicate the different initial redshifts. The blue, red and magenta markers indicate the convergence time for initial conditions $\Mg = 0.1,0.01.0.005\,M_{\rm h}(z=z_0)$, respectively, at the initial redshift indicated by the symbol type. We see that the convergence time is at most two disc orbital times.}
    \label{fig:converge}
\end{figure}

In \S\ref{sec:Results}, we rely on the assumption that the numerical solutions to equation \ref{eq:mass_conservation} converge towards a unique steady state solution, independent of the initial conditions, which can be obtained by solving the algebraic equation $\dot{M}_{\rm g,acc} = \dot{M}_{\rm trans} + \dot{M}_{\rm SF}$. In this section, we justify this assumption numerically. We define the quantity $\Delta t_{\rm SS}$ to be the time elapsed since the beginning of the integration until the deviation between the steady state solution to the true solution of eq. \ref{eq:mass_conservation} is less than $5\%$. In figure \ref{fig:converge}, we show the number of disc orbital times it takes for the solution to converge to the steady state solution, for different choices of initial conditions and different choice of initial redshifts, for different halo masses. The reason we test the convergence for different initial redshifts is that since, when presenting our results, we assume that our model is applicable only when the galaxies' host halo mass enters a certain mass range, but we still integrate the evolution of the galaxy using our model from beforehand.

We see from figure \ref{fig:converge} that the convergence time is less than two disc orbital times, for any choice of initial conditions and initial redshift. This means that from the moment the galaxy is considered disc by our model, it takes a couple of hundreds of megayears for our model to be applicable to that particular galaxy. For example, for our model with $\xi_a = 0.6$, a disc with $\log{M_{\rm h,0}}=12$ today became a disc at $z\!\sim\!2$. As a result of our integration prior to $z=2$, this galaxy had $M_{\rm g}\!\sim\!0.006\cdot M_{\rm h}(z=2)$. From our convergence test, we learn that the results of our model are applicable to this galaxy after $\sim t_{\rm orb}$ (magenta triangle), namely from $z\!\sim\!1.8$.

\bsp	
\label{lastpage}
\end{document}